\documentclass[11pt]{article}
\usepackage{graphicx,subcaption}
\usepackage{amsmath, amssymb,bbm}
\usepackage{comment,setspace}
\usepackage[margin=1in]{geometry}
\usepackage{bbm}
\usepackage{natbib}
\usepackage{booktabs}
\usepackage{url}

\title{Spatial Hyperspheric Models for Compositional Data}
\author{Michael R. Schwob$^{1,*}$, Mevin B. Hooten$^1$, Nicholas M. Calzada$^1$, Timothy H. Keitt$^2$\\
$^1$Department of Statistics and Data Sciences, The University of Texas at Austin\\
$^2$Department of Integrative Biology, The University of Texas at Austin\\
$^*$Email: schwob@utexas.edu}
\date{May 2025}

\begin{document}

\maketitle

\begin{abstract}
   Compositional observations are an increasingly prevalent data source in spatial statistics. Analysis of such data is typically done on log-ratio transformations or via Dirichlet regression. However, these approaches often make unnecessarily strong assumptions (e.g., strictly positive components, exclusively negative correlations). An alternative approach uses square-root transformed compositions and directional distributions. Such distributions naturally allow for zero-valued components and positive correlations, yet they may include support outside the non-negative orthant and are not generative for compositional data. To overcome this challenge, we truncate the elliptically symmetric angular Gaussian (ESAG) distribution to the non-negative orthant. Additionally, we propose a spatial hyperspheric regression model that contains fixed and random multivariate spatial effects. The proposed model also contains a term that can be used to propagate uncertainty that may arise from precursory stochastic models (i.e., machine learning classification). We used our model in a simulation study and for a spatial analysis of classified bioacoustic signals of the \textit{Dryobates pubescens} (downy woodpecker).
\end{abstract}


\section{Introduction}
Compositional data are non-negative and have a constant-sum constraint; they often take the form of percentages, probability vectors, or proportions \citep{aitchison1982statistical}. 
Such data have become prevalent in many scientific fields, such as genomics \citep{quinn_understanding_2018,gloor_amicompositional_2023}, microbiomics \citep{koslovsky_bayesian_2023}, business \citep{greenacre_aitchisons_2023,pal_clustering_2022}, demography \citep{martin-fernandez_interpretation_nodate}, forensics \citep{grantham_mimix_2020}, ecology \citep{nolan_comparing_2019,stewart_measuring_2022}, and environmental sciences \citep{tipton_predicting_2019}. We are interested in analyzing bioacoustic classifications, which take the form of compositional data. A composition $\mathbf{u}=(u_1,\ldots,u_d)'$ can be represented as a point on the simplex
$$\boldsymbol{\Delta}^{d-1}=\left\{(u_1,\ldots,u_d)': u_k\ge0\;\forall k, \sum_{k=1}^du_k=c\right\},$$
where $c$ is the constant in the constant-sum constraint. Without loss of generality, we assume that the composition $\mathbf{u}$ has a unit-sum constraint and set $c=1$. Most traditional statistical methods are only valid for data in Euclidean geometry. Therefore, special treatment is necessary when analyzing compositional data.
Fortunately, the growth in compositional data sources has led to recent advances in compositional data analysis (CoDA). 

Historically, compositional data were transformed to the Euclidean space through log-ratio transformations to make use of standard statistical techniques. However, this approach is problematic for compositions containing zeros \citep{aitchison1982statistical,feng2015regression}. Despite this limitation, log-ratio transformations are still the most widely used CoDA technique today \citep{greenacre2022compositional}. If compositional data contain zeros, the zeros are treated as missing and imputed or replaced with small quantities to facilitate log-ratio transformations; the underlying assumption in these cases is that the zeros are unobserved small quantities that were rounded to zero \citep{tsagris_dirichlet_2018}. However, if the value is truly zero (a structural zero), such procedures are problematic because the modification of the zero-valued components implies a change in all components due to the unit-sum constraint. 
An alternative to the log-ratio transformation is the chiPower transformation, which allows for zero-valued components at the cost of slightly violating subcompositional coherence \citep{greenacre2024chipower}.
Additionally, \cite{leininger_spatial_2013} proposed a spatial regression framework for compositional data that allows for zero-valued components while at least one component is non-zero for all observations; this is a weaker assumption but precludes some compositional data sets from analysis. 

Another common method for CoDA is Dirichlet modeling. The Dirichlet distribution is a natural choice for CoDA because its support lies on the unit simplex. Early studies relying on Dirichlet distributions were challenged by compositions containing zeros, but recent work has overcome this challenge \citep{tsagris_dirichlet_2018,koslovsky_bayesian_2023}. 
Notably, the Dirichlet distribution does not allow for positive correlations between components in a composition. 
The nested Dirichlet distribution, or the Dirichlet-tree distribution, is an extension that allows for positive correlations between components by constructing a tree of Dirichlet distributions in which components that fall under the same branch are positively correlated \citep{dennis1991hyper}. Typically, the structure of the tree is determined \textit{a priori} or is chosen via model selection techniques; the former method assumes knowledge regarding which components should be positively correlated, and the latter method is computationally expensive for a moderately large $d$ because many tree structures must be fit to the data \citep{rodriguez2008nested,tang2018phylogenetic}. 
Alternative approaches include the polynomially tilted pairwise interaction model \citep{scealy_score_2023} and the power interaction model \citep{yu2024interaction}, which aim to infer pairwise interactions between components and are fit via score matching methods for the simplex. Notably, the Dirichlet distribution is a special case of the class of power interaction models.
 
\begin{figure}
    \centering
    \begin{subfigure}{0.49\linewidth}
        \centering
        \includegraphics[width=\linewidth]{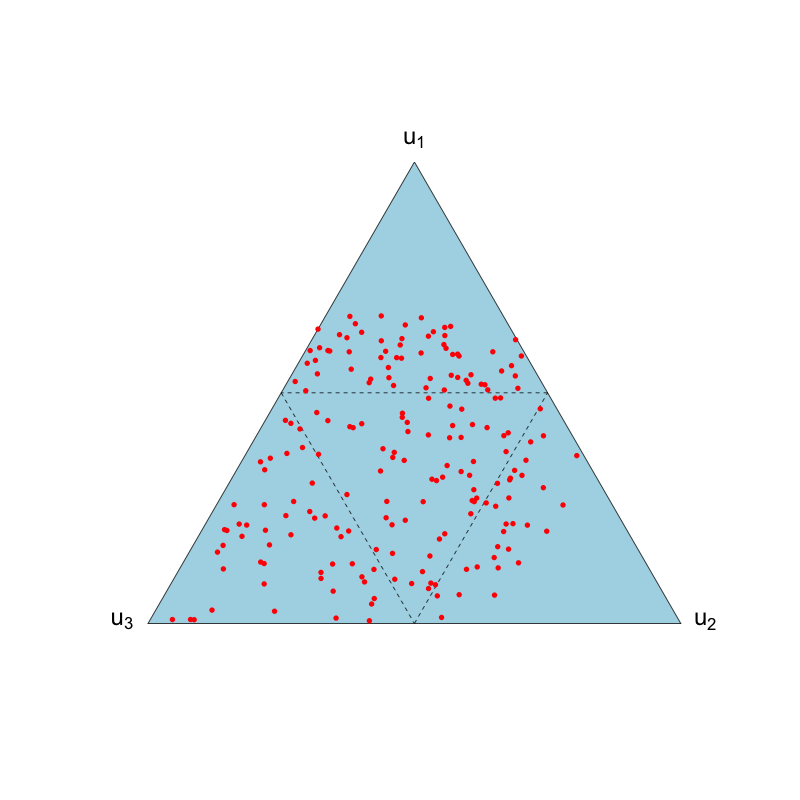}
    \end{subfigure}
    \hfill
    \begin{subfigure}{0.49\linewidth}
        \centering
        \includegraphics[width=\linewidth]{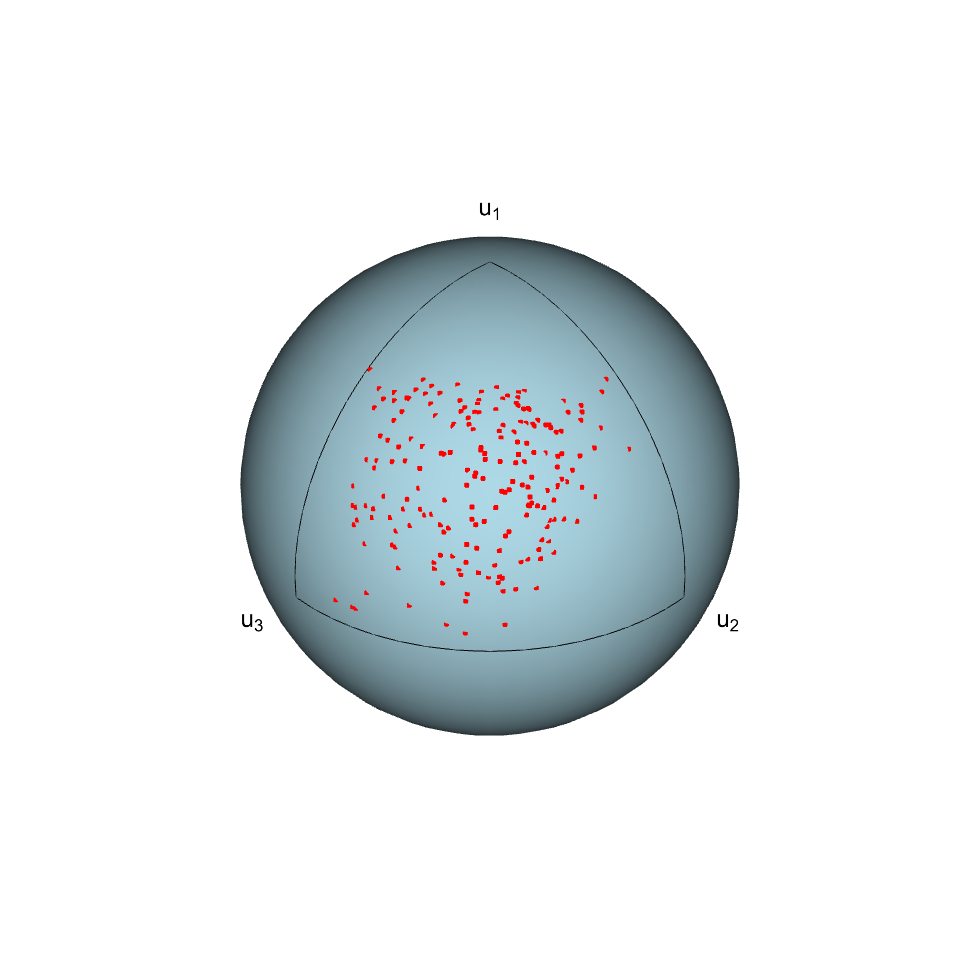}
    \end{subfigure}
    \caption{\textit{Left}: A ternary plot of $n=200$ compositions of $d=3$ dimensions.  \textit{Right}: The square-root transformed compositions on the surface of a sphere in the non-negative orthant (black lines).}
    \label{fig:sqtransformation}
\end{figure}

A promising, yet understudied approach to CoDA is the transformation of compositional data to directional data. A composition $\mathbf{u}$ is transformed to a directional data point $\mathbf{y}=(y_1,\ldots,y_d)'$  with the square-root transformation $\mathbf{y}=\sqrt{\mathbf{u}}$, where the square-root function is applied element-wise. 
The resulting data point $\mathbf{y}$ exists on the surface of the unit hypersphere $$\mathcal{S}^{d-1}=\left\{\mathbf{y}\in\mathbb{R}^d: ||\mathbf{y}||_2=1\right\}$$
in the non-negative orthant (see Figure \ref{fig:sqtransformation}), where $||\cdot||_2$ denotes the $l_2$-norm. After $\mathbf{y}$ is transformed, directional approaches may be used to analyze the data. Most distributions for directional data naturally allow for zeros and have flexible covariance structures that allow for positive and negative correlations between transformed components. Thus, directional distributions naturally overcome the challenges inherent in the log-ratio and Dirichlet-based approaches. One such distribution is the Kent distribution, or the 5-parameter Fisher-Bingham distribution. The Kent distribution is the most common distribution used to analyze directional data, despite having an analytically intractable normalizing constant for large $d$. \cite{paine_elliptically_2018} recently proposed the elliptically symmetric angular Gaussian (ESAG) distribution, which has an analytically tractable normalizing constant. Then, \cite{paine_spherical_2020} showed that the ESAG distribution provides inference similar to that obtained by the Kent distribution but at a fraction of the computational cost. 
We use the ESAG distribution to analyze transformed compositional data (i.e., bioacoustic classifications). Our approach naturally allows for zero-valued components and positive correlations between components, thereby accommodating all compositional data sets.

Analyzing the square-root transformed compositional data with conventional directional distributions is appropriate if the data are not distributed near the boundaries of the non-negative orthant \citep{scealy_regression_2011}. However, if the transformed data are distributed near the boundaries, it is likely that directional distributions will place a non-negligible amount of support outside the non-negative orthant. Thus, directional distributions are not generative for compositional data because they may generate points outside the non-negative orthant that cannot be transformed back to the simplex. In the directional statistics literature, there are generally two suggestions to remedy such boundary issues: (i) use the alternative transformation $|\mathbf{y}|=\sqrt{\mathbf{u}}$ and model $|\mathbf{y}|$ using a folded version of a directional distribution; or (ii) use a truncated directional distribution, which is ideal if the components of $\mathbf{u}$ contain a large number of zeros. 
The former remedy was implemented in \cite{scealy_fitting_2014}, where the authors used a folded Kent distribution on the transformed data; they viewed the folding assumption as a missing data problem and inferred the missing sign resulting from the folding. Although the folding approach introduces potential bias in the form of sign imputation, it was the only feasible remedy historically; truncated directional distributions were computationally prohibitive due to intractable normalizing constants. However, the ESAG distribution can be truncated at a much smaller computational cost than most directional distributions. We propose the ESAG$^+$ distribution, where the ESAG is truncated to the non-negative orthant. We then use the ESAG$^+$ distribution to analyze and predict bioacoustic classifications in a spatial setting.

One of the scientific fields generating the most compositional data is machine learning. In particular, machine learning classification is a new and readily adopted technique that results in compositional data in the form of classification (probability) vectors \citep{murphy2022probabilistic}. However, such classification output is rarely treated as a compositional data source for downstream statistical analyses. 
We analyze classification vectors and propagate the uncertainty from the machine learning classification in our downstream analysis of the compositional data.

In our downstream analysis, we use hyperspheric regression to analyze the transformed bioacoustic classifications. In general, hyperspheric regression links the mean direction of a directional distribution to a multivariate linear model. To our knowledge, we propose the first hyperspheric regression model with spatial covariates and latent spatial random effects. 
The spatial hyperspheric regression is expressed in a Bayesian hierarchical framework. With our proposed model, we conduct a spatial analysis of bioacoustic data comprising classifications of \textit{Dryobates pubescens} (downy woodpecker)  vocalizations.

The remainder of this paper is organized as follows. In Section 2, we describe the \textit{D. pubescens} vocalizations, as well as the machine learning methods that we use to classify the vocalizations to obtain compositional data. Then, we briefly review the ESAG distribution and propose the ESAG$^+$ distribution in Section 3. In Section 4, we propose our spatial hyperspheric model to analyze bioacoustic classifications and provide a method for uncertainty propagation. We present the implementation details in Section 5 and a simulation study in Section 6. In Section 7, we return to the bioacoustic compositional analysis and discuss spatial compositional prediction. Finally, we present concluding remarks in Section 8.

\section{Bioacoustic Data}

An increasingly popular source of compositional data are machine learning classification algorithms, which aim to classify an input based on its features. The output of the classification algorithm is a probability vector (i.e., a composition), where the $i$th element corresponds to the probability that the given input belongs to the $i$th class. 
We used machine learning to classify spectrograms of bioacoustic signals from the \textit{Dryobates pubescens}. The resulting acoustic compositions were used in a downstream analysis to study the relationship between the environment and the acoustic behavior of \textit{D. pubescens}.

Bioacoustic data are valuable for assessing biodiversity, habitat health, and animal behavior \citep{sueur2015ecoacoustics,keitt2021ecology}. 
In particular, avian bioacoustic signals are often used as bioindicators or phenological cues to study climate change \citep{farina2017ecoacoustics}. We analyze bioacoustic signals of the \textit{D. pubescens} because they are non-migratory and do not learn new vocalization patterns from other birds (i.e., suboscine). These characteristics allow us to focus on the relationship between the environment and the acoustic behavior of the species. 
\textit{D. pubescens} inhabit a large portion of the United States and Canada and are known to produce three distinct signals: a laugh-like whinnying call, drumming (i.e., a rapid, rhythmic pecking), and a high-pitched chirp known as a ``pik” \citep{dodenhoff2002analysis}. Both sexes produce the three distinct signals. Whinnying plays a role in the formation of pair bonds and also serves as a territorial call \citep{kilham1962reproductive,mahan1996analysis}. Drumming is a conspecific signal used to establish and defend territories, attract mates, maintain contact with mates, and signal readiness for copulation \citep{mahan1996analysis}. Piking is a response to disturbance \citep{short1982woodpeckers} and a contact vocalization associated with courtship flights \citep{kilham1962reproductive}.

We gathered \textit{D. pubescens} bioacoustic data from the Macaulay Library at the Cornell Lab of Ornithology, which is a publicly crowd-sourced repository of animal media \citep{betancourt2012teaching}. For analysis, we retained the signals that were deemed ``scientific quality” by the Macaulay Library, free of background noise, and had a three to five star rating by the Macaulay Library user community. Additionally, we focused on bioacoustic recordings from the Northeastern United States during the springs of 2020-2023. The resulting data set comprised 91 recordings, which are identified in Appendix A. 

\begin{figure}
    \centering
    \includegraphics[width=\linewidth]{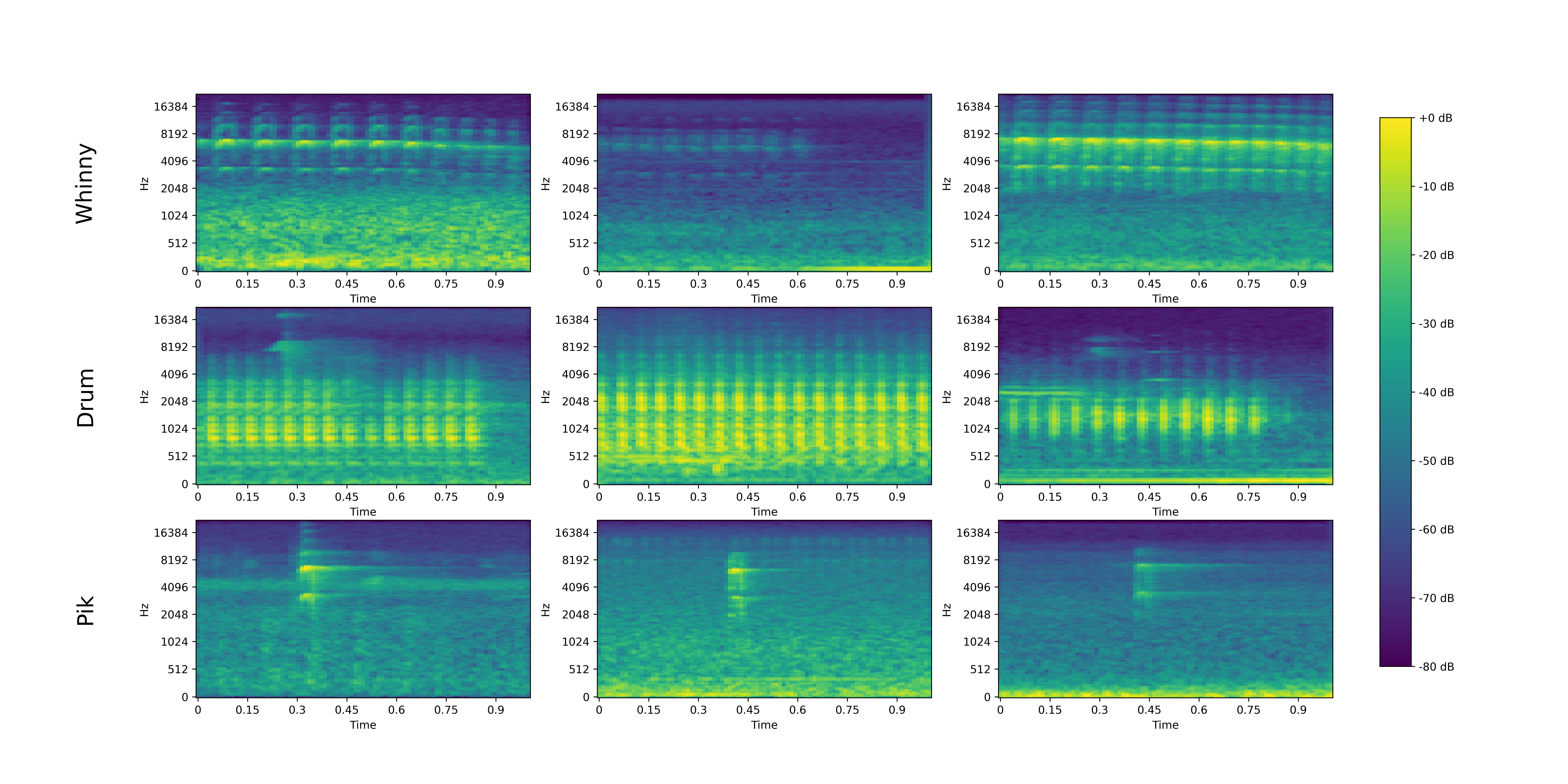}
    \caption{Three log-scaled mel spectrograms for each bioacoustic signal. These samples were randomly selected from the analyzed data.}
    \label{fig:spect_samples}
\end{figure}

We generated mel spectrograms from the recordings using the librosa Python package with a sample rate of 44.1 kHz \citep{mcfee2015librosa}. A spectrogram is a visual depiction of the spectrum of frequencies of a signal over time, where the horizontal axis represents time, the vertical axis represents frequency, and the color intensity represents amplitude. The mel spectrogram is a spectrogram where the frequencies are converted to the mel scale, which is a perceptual scale of pitches designed to represent how humans perceive sound \citep{stevens1937scale}. We converted the energy values of the mel spectrograms to a logarithmic decibel scale. Log-scaled mel spectrograms (LSMS) are often used in machine learning to capture information in a way that is more interpretable \citep{xu2018large}. We used the LSMSs as image representations of the bioacoustic signals and as the input to the machine learning classification algorithm. Figure \ref{fig:spect_samples} shows three LSMSs for each signal type of the \textit{D. pubescens}. We classified the 91 LSMSs with a two-dimensional convolutional neural network (2D-CNN) to obtain probability vectors for each recording. The 2D-CNN is commonly used to process avian acoustic signals \citep{xie2022sliding}. Details on our 2D-CNN implementation are found in Appendix B. The resulting acoustic compositional data set did not contain zero-valued components, though other machine learning classifiers may render structural zeros. Finally, we squared the elements of the prediction vector to obtain the transformed directional data for downstream analysis. 

\section{ESAG Distribution}

We consider the $d\times 1$ random vector $\mathbf{y}=\mathbf{z}/||\mathbf{z}||_2$, where $\mathbf{z}\sim\text{N}_d(\boldsymbol{\mu},\mathbf{V})$. The normalized random variable $\mathbf{y}$ follows an angular Gaussian distribution, or a projected normal distribution, with support on $\mathcal{S}^{d-1}$. The angular Gaussian distribution is denoted as $\text{AG}(\boldsymbol{\mu},\mathbf{V})$, where $\boldsymbol{\mu}$ is the mean direction and $\mathbf{V}$ controls the shape of the distribution on the surface of the hypersphere. The angular Gaussian parameters in $\boldsymbol{\mu}$ and $\mathbf{V}$ are not identifiable because $\mathbf{z}/||\mathbf{z}||_2$ and $c\mathbf{z}/||c\mathbf{z}||_2$ are equal for $c>0$ and follow the same angular Gaussian distribution despite $\mathbf{z}$ and $c\mathbf{z}$ having different means and variances when $c\ne 1$. To impose identifiability, \cite{paine_elliptically_2018} proposed the following sets of constraints: (i) $\mathbf{V}\boldsymbol{\mu}=\boldsymbol{\mu}$ and (ii) $\text{det}(\mathbf{V})=1$, where $\text{det}(\cdot)$ denotes the determinant of a matrix. These two sets of constraints lead to the construction of the ESAG distribution with probability density function
\begin{equation}\label{eq:ESAG}
    [\mathbf{y}\mid\boldsymbol{\mu},\mathbf{V}]=\frac{(2\pi)^{-(d-1)/2}}{(\mathbf{y}'\mathbf{V}^{-1}\mathbf{y})^{d/2}}\exp\left(\frac12\left(\frac{(\mathbf{y}'\boldsymbol{\mu})^2}{\mathbf{y}'\mathbf{V}^{-1}\mathbf{y}} - \boldsymbol{\mu}'\boldsymbol{\mu}\right)\right)M_{d-1}\left\{\frac{\mathbf{y}'\boldsymbol{\mu}}{(\mathbf{y}'\mathbf{V}^{-1}\mathbf{y})^{1/2}}\right\},
\end{equation}
where $M_{d-1}(t)=(2\pi)^{-1/2}\int^\infty_0x^{d-1}\exp\{-(x-t)^2/2\}dx$ and $[\cdot\mid\cdot]$ denotes the conditional probability density or mass function \citep{gelfand1990sampling}. We denote an ESAG-distributed random variable as $\mathbf{y}\sim\text{ESAG}_{d-1}(\boldsymbol{\mu},\mathbf{V})$, where $\boldsymbol{\mu}$ is the mean direction and $\mathbf{V}$ determines the orientation of dispersion. Concentration is controlled by $||\boldsymbol{\mu}||_2$ with higher values leading to more concentration and less variability. Unlike the standard angular Gaussian distribution, the ESAG has elliptical symmetry that can capture anisotropic patterns in directional data. Additionally, the normalizing constant is much easier to compute than for many existing directional distributions. Finally, simulating data from angular Gaussian distributions is relatively fast compared to most directional distributions, requiring only the normalization of a multivariate Gaussian simulation.

\subsection{An Unconstrained Parameterization}
The dimension of the parameter space for $\text{N}_{d}(\boldsymbol{\mu},\mathbf{V})$ is $d(d+3)/2$, and the sets of constraints introduced by \cite{paine_elliptically_2018} impose $d+1$ constraints. Thus, there are at most $\mathcal{P}=(d-1)(d+2)/2$ identifiable parameters for $\text{ESAG}_{d-1}(\boldsymbol{\mu},\mathbf{V})$. The original parameterization of the ESAG distribution imposed complex constraints on the model parameters. \cite{yu2024new} introduced a new parameterization of the ESAG distribution that resulted in constraint-free model parameters in $\mathbb{R}$; we provide a brief review of their parameterization in Appendix C. In particular, $\mathbf{V}$ can be specified as a function of $\boldsymbol{\mu}$ and $\boldsymbol{\gamma}$, where $\boldsymbol{\gamma}$ are constraint-free real-valued vectors. We use this new parameterization in a Bayesian hierarchical model to simplify prior specification for unconstrained parameters. In the next section, we link the mean direction $\boldsymbol{\mu}$ to a hyperspheric regression model. For $\boldsymbol{\gamma}$, we assume the following prior for each element: $\gamma_{j,k}\sim\text{N}(0,\sigma^2_\gamma)$ for $j=1,\dots,d-2$ and $k=1,\dots,j+1$.

\subsection{The Truncated ESAG Distribution}
\begin{figure}[!h]
    \centering
    \includegraphics[width=\linewidth]{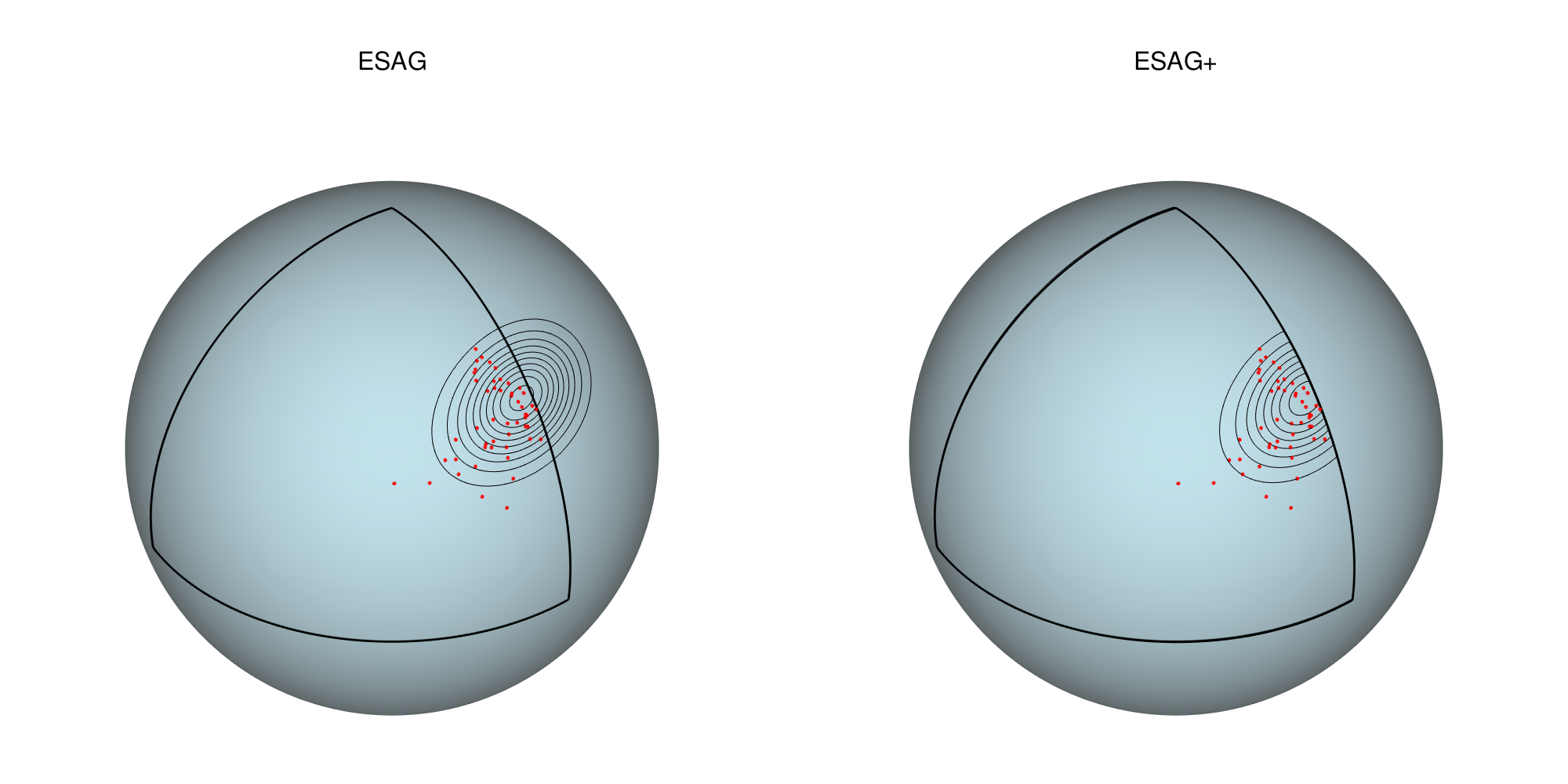}
    \caption{A comparison of the support for an ESAG$_2$ (left) and ESAG$^+_2$ (right) distribution fit to transformed compositional data (red dots) that are distributed near the boundary of the non-negative orthant.}
    \label{fig:trunc_comp}
\end{figure}

When compositional data are transformed to directional data under the square-root transformation, they lie on the surface of the non-negative orthant of the hypersphere. Conventional directional distributions may generate points outside of the non-negative orthant, which cannot be transformed back to the simplex via an element-wise squaring. Therefore, when analyzing transformed compositional data (i.e., bioacoustic classification vectors), we truncate the ESAG distribution to the non-negative orthant, denoted ESAG$^+_{d-1}$, which has probability density function
\begin{equation}\label{eq:esag_trunc}
    [\mathbf{y}\mid\boldsymbol{\mu},\mathbf{V}]^+=\frac{[\mathbf{y}\mid\boldsymbol{\mu},\mathbf{V}]\mathbbm{1}(\mathbf{y}\in\mathcal{S}^{d-1}_+)}{\int_{\mathcal{S}^{d-1}_+} [\mathbf{y}\mid\boldsymbol{\mu},\mathbf{V}]d\mathbf{y}},
\end{equation}
where $[\mathbf{y}\mid\boldsymbol{\mu},\mathbf{V}]$ is the ESAG$_{d-1}$ density function from (\ref{eq:ESAG}), $\mathcal{S}^{d-1}_+$ denotes the surface of the non-negative orthant of the unit hypersphere of $d$ dimensions, and $\mathbbm{1}(\cdot)$ is an indicator function that evaluates to 1 if the condition is satisfied and 0 otherwise. Due to the truncation, $\boldsymbol{\mu}$ is no longer the mean direction, although it is still a location parameter. See Figure \ref{fig:trunc_comp} for an example of an ESAG$_2$ distribution with support outside of the non-negative orthant and our corresponding ESAG$_2^+$ distribution. 

For many directional distributions, evaluation of the normalizing constant is computationally prohibitive. However, the normalizing constant associated with the ESAG distribution can be quickly approximated using Monte Carlo integration because simulating from the ESAG is fast. Thus, we approximate the normalizing constant for the ESAG$^+_{d-1}$ distribution with
\begin{equation}\label{eq:norm_constant}
    \int_{\mathcal{S}^{d-1}_+} [\mathbf{y}\mid\boldsymbol{\mu},\mathbf{V}]d\mathbf{y} \approx \frac{1}{M}\sum^M_{i=1}\mathbbm{1}\left(\mathbf{y}^*_i\in\mathcal{S}^{d-1}_+\right),
\end{equation}
where $\mathbf{y}^*_i$ is a random draw from $\text{ESAG}_{d-1}(\boldsymbol{\mu},\mathbf{V})$ for $i=1,\dots,M$. A quick method to simulate $\mathbf{y}_i^*$ is to simulate and normalize $\mathbf{z}_i^*\sim\text{N}_d(\boldsymbol{\mu},\mathbf{V})$. For moderately small $d$, we find that $M=5000$ Monte Carlo draws are sufficient to adequately approximate the normalizing constant. We explore the accuracy and computational efficiency of the approximation in (\ref{eq:norm_constant}) for various $d$ and $M$ in Appendix D.

Under the ESAG$^+$ distribution, the expected response $\text{E}(\mathbf{y})=\mathbf{y}^*$ has two sets of constraints: (i) $||\mathbf{y}^*||_2=1$ and (ii) $0\le y_i^* \le 1$ for $i=1,\dots,d$. These constraints are similar to the constraints imposed by a Dirichlet regression parameterized via the mean $\text{E}(\mathbf{u})$, where $\mathbf{u}$ is the compositional response: $\sum^d_{i=1}\text{E}(u_i)=1$ and $0\le \text{E}(u_i)\le 1$ for $i=1,\dots,d$. 
Under such constraints, regression parameters may not be identifiable because knowledge of $d-1$ components implies knowledge of the $d$th component.
To induce identifiability, one component is often taken as the baseline component and its regression coefficients are fixed at zero \citep{douma2019analysing}.
Baseline components (i.e., reference categories) have been used for compositional data analysis with log-ratio transformations \citep{greenacre_aitchisons_2023}, power scaling models \citep{leininger_spatial_2013}, polynomially tilted pairwise interaction models \citep{scealy_score_2023}, and Dirichlet regression \citep{douma2019analysing}. 
The constraints induced by the ESAG$^+$ distribution imply that the $d$th component of $\mathbf{y}^*$ can be obtained given the other $d-1$ components. 
Thus, we establish a baseline component when using truncated directional distributions for compositional data analysis to induce identifiability for the regression parameters. Common choices for the baseline component are the first component or the most prevalent component in the data.
For our bioacoustic compositional analysis, we selected the baseline category as ``drumming” because it was the most commonly predicted \textit{D. pubescens} signal in the classification algorithm.


\section{Hyperspheric Regression}

Hyperspheric regression is becoming a popular tool to analyze compositional data \citep{scealy_regression_2011,yu2024regression} and directional data \citep{wang2013directional,scealy_scaled_2019,paine_spherical_2020,garcia-fernandez_gaussian_2021}. The general aim is to estimate the relationship between covariates of interest and the mean direction of directional distributions; this is analogous to conventional multivariate regression approaches in Euclidean space, yet hyperspheric regression accounts for the $d$-dimensional spherical geometry in which directional data and square-root transformed compositional data reside. Recently, \cite{yu2024regression} proposed the first spherical regression framework featuring an ESAG data model. In this section, we extend their work by considering a Bayesian hierarchical hyperspheric regression model that accounts for fixed and random spatial effects. We also account for uncertainty propagation because our bioacoustic classifications arise as output from a precursory stochastic model (i.e., machine learning classification). 

\subsection{Spatial Hyperspheric Regression}

We specify a Bayesian hierarchical model that regresses directional data (i.e., transformed bioacoustic compositions) on spatial covariates and latent spatial random effects. We let $\mathbf{y}_i\sim \text{ESAG}^+_{d-1}(\boldsymbol{\mu}(\mathbf{x}_i),\mathbf{V}(\mathbf{x}_i))$ for $i=1,\dots,n$, where $n$ and $d$ denote the number of observations and the dimension of each observation, respectively. As in most first-moment regression frameworks, the location parameter $\boldsymbol{\mu}$ is specified as a function of the covariates $\mathbf{x}_i=(x_{i,1},\dots,x_{i,p})'$. Because $\mathbf{V}$ is a function of $\boldsymbol{\mu}$ in the new parameterization, we explicitly note that $\mathbf{V}(\mathbf{x}_i)$ is also a function of the spatial covariates, so the orientation of dispersion is spatially heterogeneous. 

In the standard ESAG distribution, $\boldsymbol{\mu}(\mathbf{x}_i)$ is the mean direction of the data, which lies in the non-negative orthant if the directional data are transformed compositional data. Therefore, we use the softplus link function to ensure that all elements of $\boldsymbol{\mu}(\mathbf{x}_i)$ are non-negative:
\begin{equation}
    \boldsymbol{\mu}(\mathbf{x}_i)=\text{softplus}(\mathbf{h}(\mathbf{x}_i)),
\end{equation}
where $\text{softplus}(x)=\log(1+e^x)$ is applied element-wise and $\mathbf{h}(\mathbf{x}_i)$ takes the form of a multivariate linear regression model. We use the softplus link function instead of the more conventional log-link function because we found that, as the range of concentration (i.e, $\{||\boldsymbol{\mu}(\mathbf{x}_i)||_2\}_{i=1}^n$) expands, mixing gets worse. In particular, the quasi-additive nature of the softplus function generally leads to better mixing than the multiplicative nature of the log-link function \citep{wiemann2024using}. Additionally, the softplus link function allows for a near additive interpretation of the relationship between the covariates and $\boldsymbol{\mu}(\mathbf{x}_i)$, which is generally more intuitive.

For our spatial hyperspheric regression model, we specify
\begin{equation}
    \boldsymbol{\mu}(\mathbf{x}_i)=\text{softplus}(\mathbf{Bx}_i+\boldsymbol{\eta}(\mathbf{s}_i)),
\end{equation}
where $\mathbf{B}$ is a $d\times p$ matrix of regression coefficients and $\boldsymbol{\eta}(\mathbf{s}_i)$ are the $d\times 1$ latent spatial random effects at the location of the $i$th observation, $\mathbf{s}_i$. We denote the $j$th row of $\mathbf{B}$ as $\mathbf{B}_{j,\cdot}=\boldsymbol{\beta}_j'=(\beta_{j,1},\dots,\beta_{j,p})$ and assume a multivariate Gaussian prior for each row: $\boldsymbol{\beta}_j\sim\text{N}(\boldsymbol{0}_p, \sigma^2_\beta\mathbf{I}_p)$.
The latent effects $\boldsymbol{\eta}(\mathbf{s}_i)$ suggest that each component (i.e., bioacoustic signal) has its own spatial field.

Perhaps the simplest approach to modeling $\boldsymbol{\eta}(\mathbf{s}_i)$ is to assume that the latent random effects comprise $d$ independent spatial fields. However, we assume that the spatial fields among the components are dependent and model $\boldsymbol{\eta}(\mathbf{s}_i)$ with a multivariate spatial process. Ideally, one would have knowledge about how one component may depend on another; if so, a conditional multivariate spatial process may be specified \citep{cressie2015statistics}. We assume a lack of such knowledge and specify $\boldsymbol{\eta}(\mathbf{s}_i)$ jointly. 
We model the multivariate spatial process with a linear model of coregionalization (LMC), which provides a flexible framework that can capture complex dependence structures \citep{schmidt2003bayesian,gelfand2010}. 
In particular, we specify the matrix-variate LMC prior $\mathbf{H}\sim\text{LMC}(\mathbf{C},\mathbf{R}_{j=1}^d)$ with density function
\begin{equation}\label{eq:LMCdens}
    p(\mathbf{H}\mid\mathbf{C},\mathbf{R}_{j=1}^d)=\frac{\exp\left(-\frac{1}{2}\sum^d_{j=1}\mathbf{c}_j^{-1}\mathbf{H}\mathbf{R}_j^{-1}\mathbf{H}'(\mathbf{c}_j^{-1})'\right)}{(2\pi)^{nd/2}|\text{det}(\mathbf{C})|^n \prod^d_{j=1}\text{det}(\mathbf{R}_j)^{1/2}},
\end{equation}
where $\mathbf{H}$ is a $d\times n$ matrix with columns $\boldsymbol{\eta}(\mathbf{s}_1),\dots,\boldsymbol{\eta}(\mathbf{s}_n)$, the $\{\mathbf{R}_j\}$ are spatial correlation matrices obtained by applying the correlation functions $\rho(|\mathbf{s}_i-\mathbf{s}_{k}|;\phi_j)$ to all pairs of locations, $\mathbf{c}_j$ is the $j$th column of the coregionalization coefficients matrix $\mathbf{C}$, and $\mathbf{c}_j^{-1}$ is the $j$th row of $\mathbf{C}^{-1}$.
If observations are co-located, there are $m<n$ distinct locations; without loss of generality, we let $m=n$.
\cite{alie2024computational} show that (\ref{eq:LMCdens}) is equivalent to
$$\text{vec}(\mathbf{H})=\boldsymbol{\eta} \sim \text{N}_{dn}\left(\boldsymbol{0},\sum^d_{j=1}\mathbf{R}_j\otimes \mathbf{c}_j\mathbf{c}_j'\right),$$
where
\begin{equation}\label{eq:eta}
    \boldsymbol{\eta}=(\eta(\mathbf{s}_1)_1,\dots,\eta(\mathbf{s}_1)_d,\dots,\eta(\mathbf{s}_n)_1,\dots,\eta(\mathbf{s}_n)_d)'
\end{equation}
is the $dn\times 1$ stacked vector of latent spatial random effects and
$\otimes$ denotes the Kronecker product.
We use the matrix-variate parameterization of the LMC due to its computational savings \citep{zhang2022spatial}. In particular, computation can be executed directly on $\mathbf{C}$ and $\{\mathbf{R}_j\}$ rather than computing the determinant and inverse of the $nd\times nd$ quadratic product under the vectorized form \citep{kyzyurova2017uncertainty}.

The coregionalization matrix $\mathbf{CC}'$ captures the relationships between spatially correlated variables and decomposes the overall variance into contributions from individual variables and their covariances \citep{schmidt2003bayesian}. A common approach to modeling $\mathbf{C}$ is to specify a lower triangular parameterization to induce identifiability. However, such a constraint imposes an asymmetry on the components of the model, which is limiting and often unjustifiable \citep{alie2024computational}. Although $\mathbf{C}$ is not identifiable, the positive definite matrix $\mathbf{CC}'$ is identifiable, which is sufficient for the identifiability of the multivariate spatial process $\boldsymbol{\eta}$. We assume a lack of knowledge about across-component covariance and specify independent standard normal priors for each element of $\mathbf{C}$.

We construct the $n\times n$ spatial correlation matrices by specifying the Matérn 3/2 correlation function
$$\rho(|\mathbf{s}_i-\mathbf{s}_k|; \phi_j)=\left(1+\frac{\sqrt{3}|\mathbf{s}_i-\mathbf{s}_k|}{\phi_j}\right)\exp\left(-\frac{\sqrt{3}|\mathbf{s}_i-\mathbf{s}_k|}{\phi_j}\right)$$
for $j=1,\dots,d$, where $|\mathbf{s}_i-\mathbf{s}_k|$ is the Euclidean distance between locations $\mathbf{s}_i$ and $\mathbf{s}_k$. We model spatial dependency with the Matérn 3/2 correlation function because it offers flexibility in modeling smoothness, providing a balance between smooth functions (such as Gaussian decay) and rougher functions (such as exponential decay). Additionally, the moderate complexity of the Matérn 3/2 function reduces the risk of overfitting compared to higher-order Matérn correlation functions \citep{hoeting2006model}. We value the flexibility in modeling smoothness because we assume a lack of knowledge about the smoothness of bioacoustic signal classification. In other applications, the selection of $\rho(|\mathbf{s}_i-\mathbf{s}_{k}|;\phi)$ can be aided by domain expertise, exploratory data analysis, and model selection. For our bioacoustic analysis, we assign independent informative priors $\phi_j\sim\text{Gamma}(62,0.04)$, which correspond to practical ranges at distances between 0.01 and half of the maximum observed distance in our bioacoustic data. We use informative priors because the spatial range parameters cannot be consistently estimated along with the marginal variances contained in $\mathbf{C}$ \citep{zhang2004inconsistent}.

\subsection{Uncertainty Propagation}

We consider the case where the compositional or directional data are obtained as output from another model; in our bioacoustic application, the compositions are obtained from a machine learning classifier, which inherently introduces uncertainty.
When output from model $A$ is treated as input to model $B$, it is typically desirable to account for any uncertainty associated with the output of model $A$ in model $B$ \citep{draper1995assessment}. We account for such uncertainty by introducing an additive term to the hyperspheric regression model:
\begin{equation}
    \boldsymbol{\mu}(\mathbf{x}_i,\mathbf{z}_i)=\text{softplus}(\mathbf{Bx}_i+\boldsymbol{\eta}(\mathbf{s}_i) + \mathbf{A}\mathbf{z}_i),
    \label{eq:softplus}
\end{equation}
where $\mathbf{z}_i$ is a $q$-dimensional vector of covariates used to quantify uncertainty in model $A$ predictions, $\mathbf{A}=\mathbf{1}_d\otimes \boldsymbol{\alpha}'$ is a $d\times q$ matrix, $\mathbf{1}_d$ is a $d$-dimensional vector of ones, and $\boldsymbol{\alpha}=(\alpha_1,\dots,\alpha_q)'$ are regression coefficients for $\{\mathbf{z}_i\}_{i=1}^n$. The role of $\mathbf{Az}_i$ is to attenuate the concentration of the data model at the observation level. If an observation was predicted by model $A$ with a high level of uncertainty, then $\mathbf{Az}_i$ is expected to be smaller for that observation (i.e., decreasing the concentration). Similarly, if an observation was predicted by model $A$ with a high level of certainty, then $\mathbf{Az}_i$ is expected to be larger (i.e., increasing the concentration). The uncertainty propagation term $\mathbf{Az}_i$ must be included in the specification for $\boldsymbol{\mu}$ because its $l_2$-norm specifies the concentration. However, we do not want $\mathbf{Az}_i$ to contribute to the learning of the normalized mean direction (i.e., to affect the point at which the mean direction intersects the surface of the non-negative orthant). Thus, we specify $\mathbf{A}$ to have identical rows comprising $\boldsymbol{\alpha}$. We consider the prior $\boldsymbol{\alpha}\sim\text{N}(\boldsymbol{0}_q,\sigma^2_\alpha\mathbf{I}_q)$ in our Bayesian hierarchical model. The complete model specification is outlined in Appendix E.

\section{Implementation}

We fit the proposed Bayesian hierarchical hyperspheric regression model using an MCMC algorithm. We used Metropolis-Hastings updates for $\boldsymbol{\alpha}$, $\{\boldsymbol{\beta}_j\}_{j=1}^d$, $\boldsymbol{\gamma}$, $\{\phi_j\}_{j=1}^d$, and the elements of $\mathbf{C}$. For $\boldsymbol{\alpha}$, we used a multivariate symmetric random-walk proposal: $\boldsymbol{\alpha}^{(*)}\sim\text{N}(\boldsymbol{\alpha}^{(k-1)},\sigma^2_\alpha\mathbf{I})$, where $(*)$ denotes the proposed values and $(k-1)$ denotes the values on the previous MCMC iteration. We also used a symmetric random-walk proposal for each row of $\mathbf{B}$, each element of $\boldsymbol{\gamma}$, and each element of $\mathbf{C}$: 
\begin{align*}
    \boldsymbol{\beta}_j^{(*)} &\sim \text{N}(\boldsymbol{\beta}_j^{(k-1)},\sigma^2_\beta\mathbf{I}), \quad j=1,\dots,d,\\
    \gamma_{l,i}^{(*)} &\sim \text{N}(\gamma_{l,i}^{(k-1)},\sigma^2_\gamma), \quad l=1,\dots,d-2, \; i=1,\dots,l+1,\\
    C_{ij}^{(*)} &\sim \text{N}(C_{ij}^{(k-1)},\sigma^2_c), \quad i=1,\dots,d, \; j=1,\dots,d,
\end{align*}
where we only accepted the proposal $C_{ij}^{(*)}$ if the resulting coregionalization matrix $\mathbf{CC}'$ was positive definite \citep{gelfand1992bayesian}.
We used the proposal $\phi_j^{(*)}\sim\text{N}(\phi_j^{(k-1)},\sigma^2_\phi)_0^\infty$ with a random-walk truncated below 0 to ensure the spatial range was non-negative. 
Finally, we updated the $nd$-dimensional vector $\boldsymbol{\eta}$ with an elliptical slice sampler, which provided significantly better mixing than Metropolis algorithms for moderately high dimensions \citep{murray2010elliptical}. Then, we constructed $\mathbf{H}$ from $\boldsymbol{\eta}$, which was used to evaluate the matrix-variate LMC density in (\ref{eq:LMCdens}) during the updates for $\{\phi_j\}_{j=1}^d$ and the elements of $\mathbf{C}$.

For computational efficiency, we implemented the MCMC algorithm in Julia and used static arrays to significantly decrease computation costs \citep{bezanson2017julia}. We also evaluated the $M_{d-1}(\cdot)$ term of the ESAG and ESAG$^+$ densities using adaptive Gauss–Kronrod quadrature, which balanced computational speed and accuracy \citep{quadgk}. Finally, we approximated the normalizing constant for the ESAG$^+$ density in (\ref{eq:norm_constant}) using parallelized Monte Carlo integration. 
In both the simulation study and the bioacoustic analysis, we ran the MCMC algorithm until the $\hat{R}$ convergence diagnostic was less than 1.01 for all parameters in the model; the spatial range parameters $\{\phi_j\}$ required the most iterations to satisfy this condition \citep{zhang2004inconsistent}. We ran our MCMC algorithm for 300000 iterations and 400000 iterations for the simulation study and bioacoustic analysis, respectively. The computational costs of our model are provided for the simulation study and the bioacoustic application in the respective sections.

\subsection{Compositional Inference}

If the original data are compositional (i.e., bioacoustic classifications), there may be interest in predicting a new composition given a set of covariates. \cite{yu2024new} suggested that predicting a composition is equivalent to estimating $\text{E}(\mathbf{y}^2)=\boldsymbol{\Xi}^2\text{E}(\mathbf{K}^2)$, where $\boldsymbol{\Xi}=(\boldsymbol{\xi}_d 
\,\,\, \boldsymbol{\xi}_{d-1}\,\, \dots \,\, \boldsymbol{\xi}_1)$, $\mathbf{K}=\boldsymbol{\Xi}'\mathbf{y}$, and the square is applied element-wise. The $\{\boldsymbol{\xi}\}_{i=1}^d$ are defined in Appendix C as the orthonormal eigenvectors of $\mathbf{V}$ and can be obtained deterministically given $\boldsymbol{\mu}$ and $\boldsymbol{\gamma}$. A Monte Carlo estimator for $\text{E}(\mathbf{y}^2)$ is
$$\widehat{\text{E}(\mathbf{y}^2)}=\widehat{\boldsymbol{\Xi}}^2\sum^M_{i=1}\widehat{\mathbf{K}}^2_i/M,$$
where $M$ is the number of Monte Carlo simulations, $\widehat{\mathbf{K}}_i=\widehat{\boldsymbol{\Xi}}'\mathbf{y}_i$, and $\widehat{\boldsymbol{\Xi}}$ is the posterior mean of $\boldsymbol{\Xi}$. An alternative estimator approximates $\text{E}(\mathbf{y}^2)$ via Monte Carlo integration:
\begin{equation}\label{eq:estComp}
    \int_{\mathcal{S}_{d-1}^+}\mathbf{y}^2\cdot[\mathbf{y}\mid\boldsymbol{\mu},\mathbf{V}]^+d\mathbf{y} \approx \frac{1}{M}\sum^M_{i=1}\mathbf{y}_i^2\cdot[\mathbf{y}_i\mid\hat{\boldsymbol{\mu}},\hat{\mathbf{V}}]^+,
\end{equation}
where $[\mathbf{y}_i\mid\boldsymbol{\mu},\mathbf{V}]^+$ is the probability density function for the ESAG$^+$ distribution in (\ref{eq:esag_trunc}), $\hat{\boldsymbol{\mu}}$ is the posterior mean of $\boldsymbol{\mu}$, and $\hat{\mathbf{V}}=f(\hat{\boldsymbol{\mu}},\hat{\boldsymbol{\gamma}})$, where $\hat{\boldsymbol{\gamma}}$ is the posterior mean of $\boldsymbol{\gamma}$ and $f(\cdot)$ is specified in Appendix C.
Uniformly random samples from $\mathcal{S}_{d-1}^+$ can be obtained by computing $\mathbf{y}_i=|\mathbf{z}_i|/||\mathbf{z}_i||_2$, where $|\cdot|$ is applied element-wise and $\mathbf{z}_i\sim\text{N}(\boldsymbol{0}_d,\mathbf{I}_d)$ for $i=1,\dots,M$.

\section{Simulation Study}

We simulated directional data from our Bayesian hierarchical model and aimed to recover the true parameter values. First, we uniformly sampled $n=100$ observations from the study domain $\mathcal{S}=[0,1]\times[0,1].$ The covariates for each observation were fully specified by their spatial location: $\mathbf{x}(\mathbf{s}_i)=(1,x(\mathbf{s}_i)_1,x(\mathbf{s}_i)_2)'$ for $i=1,\dots,100$, where $x(\mathbf{s}_i)_1=|s_{i,1}-0.5|^{1.2}$ and $x(\mathbf{s}_i)_2=||\mathbf{s}_i||_2$. Then, we standardized the environmental covariates to be in $[1,2]$, which is more comparable in scale with directional data \citep{scealy_scaled_2019}. We set the first of three dimensions as the baseline component and specified
$$\mathbf{B}=\begin{bmatrix}
1.0 & 0.0 & 0.0 \\
1.25 & 1.15 & 0.25 \\
1.1 & 0.55 & 1.65 
\end{bmatrix} \quad , \quad \mathbf{C}=\begin{bmatrix}
1.46 & -0.91 & 0.43 \\
1.15 & -0.60 & 1.35 \\
1.18 & 1.54 & 1.09 
\end{bmatrix},$$
$\boldsymbol{\phi}=(0.12, 0.15, 0.19)'$, $\alpha=-0.31$, and $\boldsymbol{\gamma}=(0.728,0.346)'$. 
Then, we simulated $\boldsymbol{\eta}$, computed $\boldsymbol{\mu}(\mathbf{x}_i)$ and $\mathbf{V}(\mathbf{x}_i)$, and simulated the directional response $\mathbf{y}_i$ from the ESAG$^+$ data model. We specified the following hyperparameters: $\sigma^2_\alpha=10$, $\sigma^2_\beta=10$, and $\sigma^2_\gamma=10$. We specified the hyperparameters in $\phi_j\sim\text{Gamma}(\alpha_\phi,\theta_\phi)$ for $j=1,\dots,d$ such that 95\% of the prior support was less than half of the observed maximum distance with a standard deviation of 0.03.

We fit two versions of our Bayesian hierarchical model to the simulated data: (i) with an ESAG$^+$ data model with the first component set as the baseline and (ii) with an ESAG data model (which did not need a baseline component). The models were fit to the simulated data with an MCMC algorithm that ran for 300000 iterations with a burn-in of 60000 iterations. The ESAG model and ESAG$^+$ model performed 41 it/s and 6 it/s, respectively, on a 3.2 GHz processor with 64 GB of RAM. For the ESAG$^+$ model, the effective sample sizes for $\{\phi_j\}$ were 1725.213, 1728.566, and 1694.102. The average effective sample size for $\boldsymbol{\eta}$ was 1693.296 with a minimum of 1620.348 and a maximum of 1735.055. The effective sample sizes under the ESAG model were comparable.

\begin{figure}
    \centering
    \includegraphics[width=\linewidth]{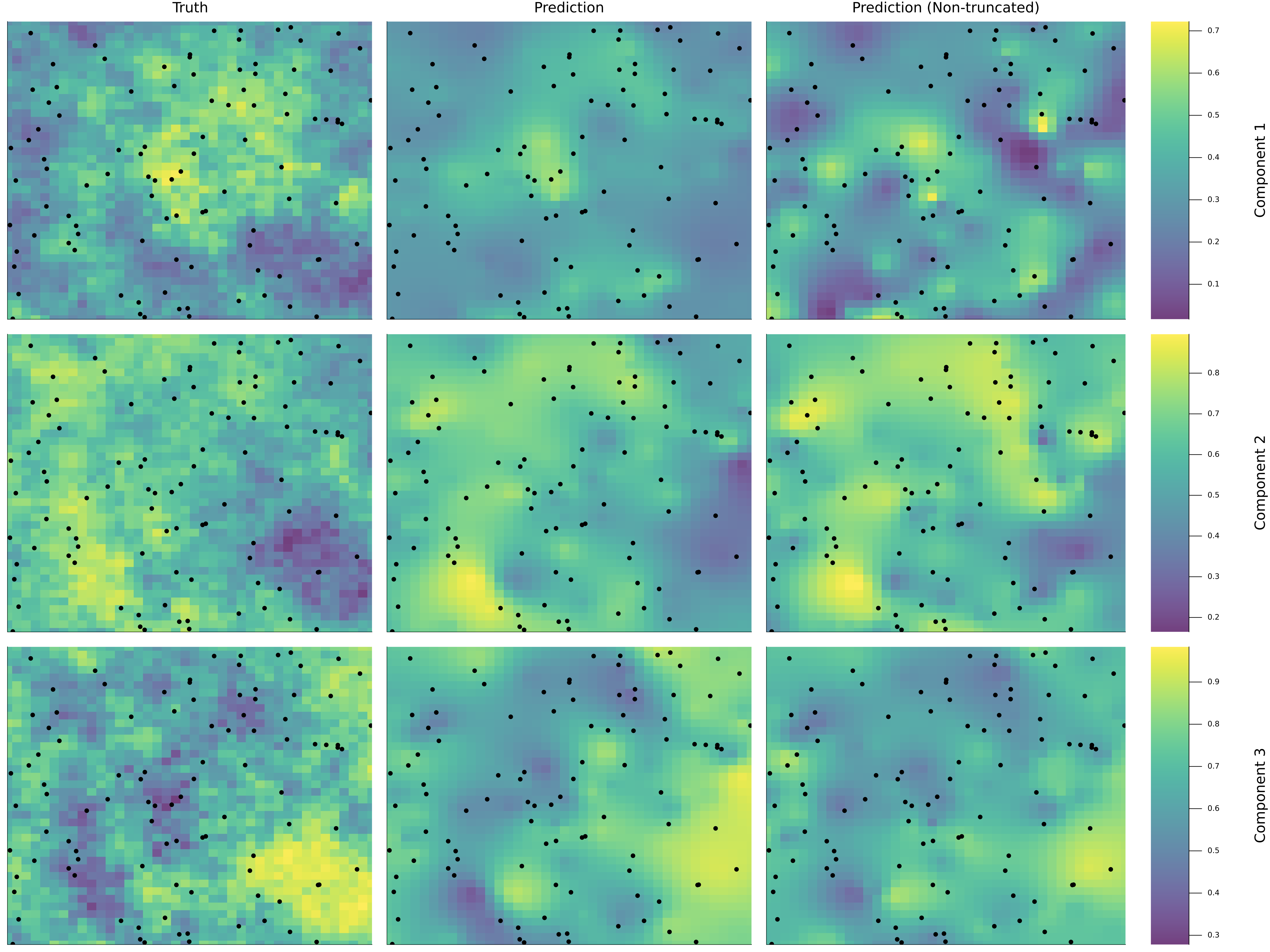}
    \caption{The true normalized mean direction (left column) and the posterior normalized mean directions from the ESAG$^+$ model (center column) and ESAG model (right column). The black dots denote the observed locations.}
    \label{fig:nmd_compare}
\end{figure}

Both models captured the true values for $\alpha$ and $\boldsymbol{\gamma}$ in their 95\% credible intervals. Only the ESAG$^+$ model captured $\boldsymbol{\phi}$ in 95\% credible intervals. However, the spatial range parameters are challenging to estimate for multivariate spatial processes \citep{gelfand2004nonstationary,zhang2004inconsistent}. The true regression coefficients $\mathbf{B}$ were recovered by both models, though the posterior means were closest under the ESAG$^+$ model. For both models, we estimated the normalized mean direction $\text{E}(\mathbf{y})$ across a fine grid spanning the study domain. The posterior normalized mean directions are compared to the true normalized mean directions in Figure \ref{fig:nmd_compare}, which visually indicates that the ESAG$^+$ model better recovered the truth as expected. 

For quantitative model comparison, we used the logarithmic score function (logS) and the $\chi^2$ measure of compositional distance (CSD). The logS is commonly used to score multivariate predictions because it evaluates the mean and variance of the marginal distribution and the correlation structure of the prediction \citep{bjerregaard2021introduction}. The mean logS were 3.265 and 1.621 for the ESAG and ESAG$^+$ models, respectively, indicating that the ESAG$^+$ model provided a better fit to the data. Out of a set of compositional distance measures that were scale and permutation invariant and allowed for zero-valued components, \cite{stewart_approach_2017} found that the CSD was the measure that violated subcompositional coherence the least. We used CSD to compare the squared observations with the predicted compositions of the competing models. In particular, we used (\ref{eq:estComp}) to predict the composition for the ESAG$^+$ model and a similar Monte Carlo integration containing the non-truncated density to predict the composition for the ESAG model. The mean CSDs were 0.603 and 0.542 for the ESAG and ESAG$^+$ models, respectively, further suggesting that the ESAG$^+$ model provided the better fit to the data.

\section{Bioacoustic Compositional Analysis}

We fit our Bayesian hierarchical model in Appendix E to the \textit{D. pubescens} bioacoustic data. We considered two environmental covariates that we hypothesized influence \textit{D. pubescens} acoustic behavior: temperature and precipitation \citep{dodenhoff2002analysis}. In particular, we collected the average temperature in degrees Celsius ($\mathbf{x}_1$) and precipitation in millimeters ($\mathbf{x}_2$) at the observed locations across the spring (March-May) in which the bioacoustic signals were recorded. Similar to \cite{scealy_scaled_2019}, we transformed the covariates with $x^*_{i,k}=(x_{i,k}-x_{(1),k})/(x_{(n),k}-x_{(1),k})+1$ for $i=1,\dots,n$, where $x_{(1),k}$ and $x_{(n),k}$ are the minimum and maximum ordered statistics for the $k$th covariate, respectively. The transformed covariates were more comparable in scale with the response of a unit Euclidean norm. The final design matrix was $\mathbf{X}=[\mathbf{1}_n \;\; \mathbf{x}_1^* \;\; \mathbf{x}^*_2]$. 

For uncertainty propagation, we considered the Macaulay Library user community ratings to be the auxiliary information $\mathbf{z}_i=(z_{i,1},z_{i,2})'$, where $z_{i,1}$ and $z_{i,2}$ are indicator covariates that equal 1 if the community rating is four stars or three stars, respectively. We used community ratings as the auxiliary information because lower quality data were assumed to have more noise and lead to less confident (more variable) predictions in the machine learning classification.

We specified the following hyperparameters in our model: $\sigma^2_\alpha=10$, $\sigma^2_\beta=10$, and $\sigma^2_\gamma=10$. Such variances are vague for an ESAG regression because the magnitude of the regression parameters contribute quadratically in $d$ to the $l_2$-norm of $\boldsymbol{\mu}$, which is typically small. For the spatial range parameters, we assigned independent informative priors $\phi_j\sim\text{Gamma}(62,0.04)$, which corresponded to practical ranges of avian soundscapes at distances between 0.01 and half of the maximum observed distance in the data \citep{farina2017ecoacoustics}. Finally, we specified random-walk proposals with variances of $0.1$, $0.1$, $0.1$, $0.01$, and $0.5$ for $\boldsymbol{\alpha}$, $\boldsymbol{\gamma}$, $\mathbf{C}$, $\boldsymbol{\phi}$, and $\boldsymbol{\beta}$, respectively. We implemented the hierarchical model in an MCMC algorithm in Julia, which we ran for 400000 iterations with a burn-in of 160000 iterations. We used $M=5000$ Monte Carlo draws for approximating the normalizing constant in (\ref{eq:norm_constant}). The MCMC algorithm performed 8 it/s on a 3.2 GHz processor with 64 GB of RAM. The effective sample sizes for $\{\phi_j\}$ were 1515.658, 1506.136, and 1508.484. The average effective sample size for $\boldsymbol{\eta}$ was 2321.221 with a minimum of 1571.198 and a maximum of 3169.788.

The posterior mean for $\boldsymbol{\alpha}$ was $\widehat{\boldsymbol{\alpha}}=(-0.62516,-1.02445)'$. Thus, lower community ratings were estimated to decrease the concentration of $\boldsymbol{\mu}$, thereby propagating more uncertainty in the data model; this was expected because it was assumed that community ratings were negatively correlated with noise. 
The posterior mean for $\mathbf{B}$ was
$$ \widehat{\mathbf{B}}=\begin{bmatrix}
-1.05524 & -1.11464 & -0.370964 \\
-5.17297 & 0.0 & 0.0 \\
0.847705 & -0.396902 & -2.22877 
\end{bmatrix},$$ 
where the rows correspond to the environmental effects of identifying whinnying, drumming, and piking, respectively. 
Inference on the environmental effects must be made relative to the baseline category (i.e., drumming).
The magnitude of the intercept for the drumming category was quite large to offset the exclusively negative relative relationships that the other signals have with the environmental covariates. We did not draw inference from $\widehat{\mathbf{B}}$ directly because $\boldsymbol{\mu}$ is not the mean direction under the ESAG$^+$ model. Rather, we predicted classifications of \textit{D. pubescens} bioacoustic signals across the study domain and compared the predictions to maps of the environmental covariates.


\subsection{Spatial Compositional Prediction}
\begin{figure}
    \centering
    \includegraphics[width=\linewidth]{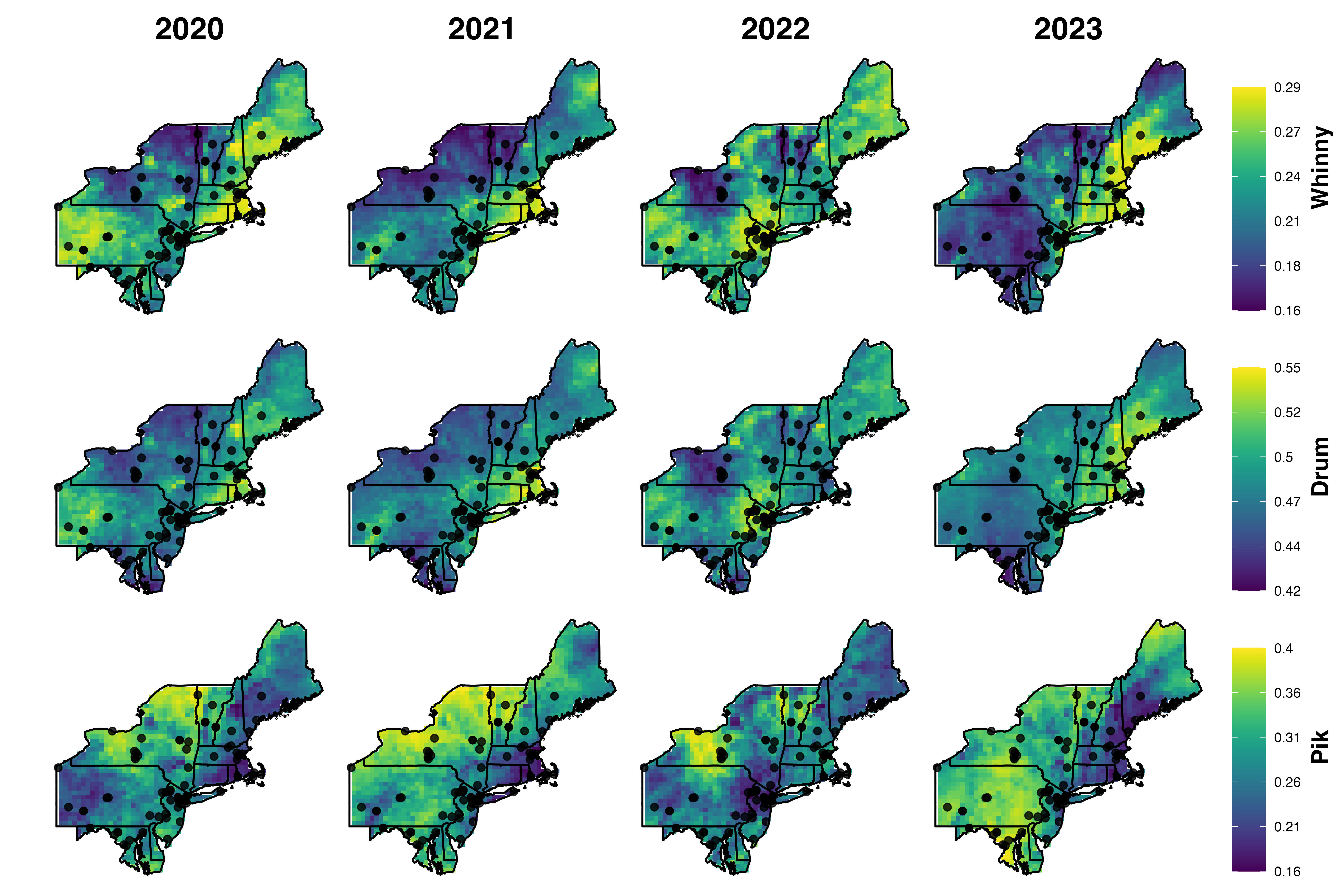}
    \caption{Predicted \textit{Dryobates pubescens} bioacoustic signal compositions for the Northeastern United States during spring from 2020-2023. Black dots denote observed locations.}
    \label{fig:pred-comp}
\end{figure}

We predicted the bioacoustic classification compositions across the Northeastern United States for each spring during the study period. To do this, we collected and standardized the average temperature ($^\circ$C) and precipitation (mm) at a new location $\mathbf{s}^*$ for each spring. Then, we estimated $\boldsymbol{\mu}(\mathbf{s}^*)$ with the posterior mean
$$\text{E}(\boldsymbol{\mu}(\mathbf{s}^*)\mid \mathbf{Y})=\widehat{\boldsymbol{\mu}}(\mathbf{s}^*)\approx\frac{1}{K}\sum^{K}_{k=1}\text{softplus}(\mathbf{B}^{(k)}\mathbf{x}(\mathbf{s}^*) + \boldsymbol{\eta}^{(k)}(\mathbf{s}^*))$$
for each spring, where the $(k)$ superscript denotes the $k$th MCMC iteration and $\boldsymbol{\eta}^{(k)}(\mathbf{s}^*)$ maximizes
$$[\boldsymbol{\eta}(\mathbf{s}^*)\mid \boldsymbol{\eta}^{(k)}]=\int\int [\boldsymbol{\eta}(\mathbf{s}^*),\boldsymbol{\phi}^{(k)},\mathbf{C}^{(k)}\mid \boldsymbol{\eta}^{(k)}]d\boldsymbol{\phi}^{(k)}d\mathbf{C}^{(k)}.$$ 
Finally, we estimated the bioacoustic classification composition at $\mathbf{s}^*$ using (\ref{eq:estComp}) with $\widehat{\boldsymbol{\mu}}(\mathbf{s}^*)$ and $\widehat{\mathbf{V}}(\mathbf{s}^*)=f(\widehat{\boldsymbol{\mu}}(\mathbf{s}^*),\widehat{\boldsymbol{\gamma}})$, where $\widehat{\boldsymbol{\gamma}}$ is the posterior mean of $\boldsymbol{\gamma}$. 
We predicted acoustic compositions for each $\mathbf{s}^*\in\mathcal{S}$, where $\mathcal{S}$ is a fine grid spanning the study domain.
The predicted acoustic compositions are plotted in Figure \ref{fig:pred-comp} for the springs of 2020-2023. Such predictions assume that the environmental effects on \textit{D. pubescens} bioacoustic signal identification are time invariant. That is, the classification of \textit{D. pubescens} acoustic behavior does not change in time and is only influenced by the local environment. 

\subsection{Results}

\begin{figure}
    \centering
    \includegraphics[width=\textwidth]{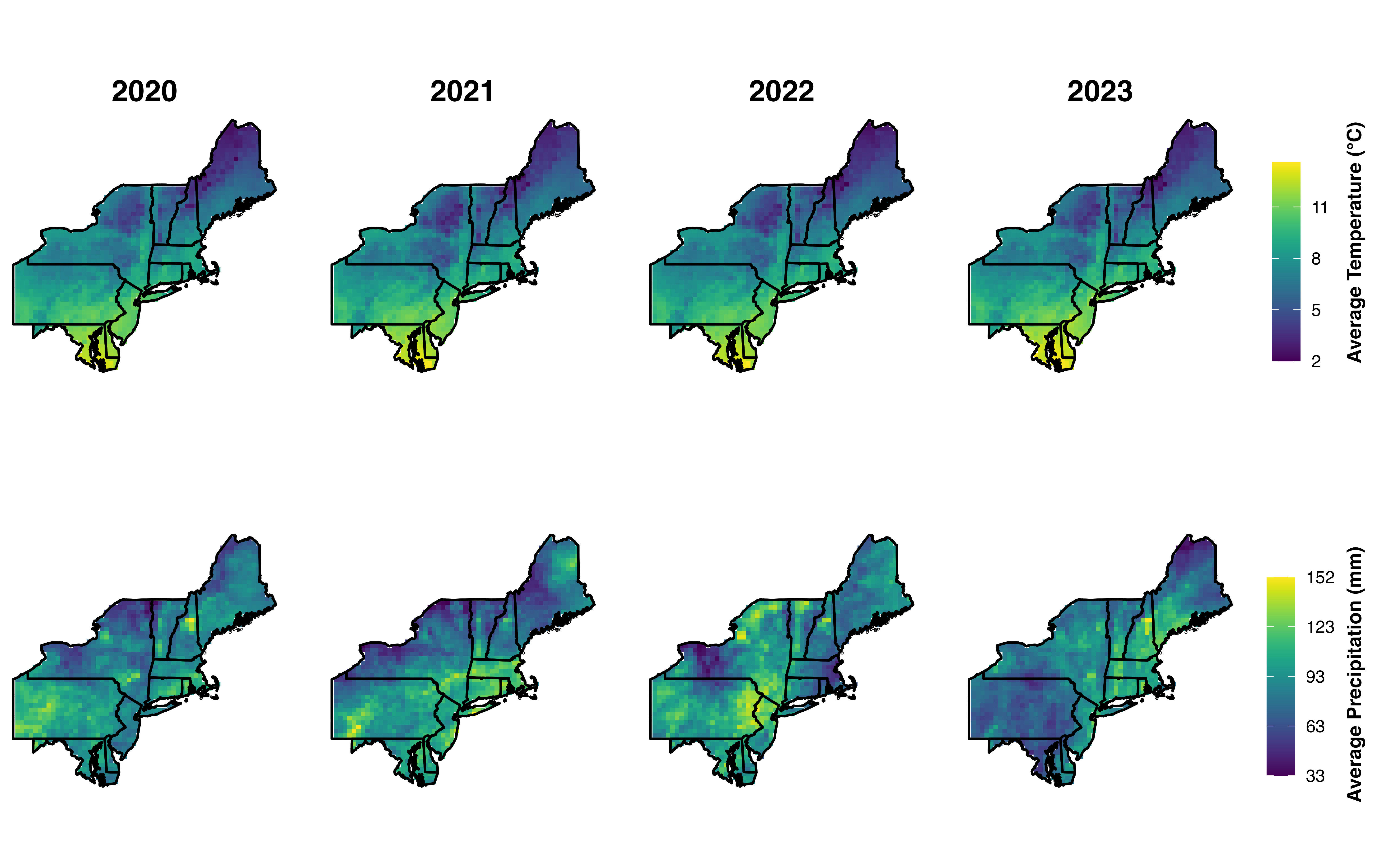}
    \caption{Average spring temperature ($^\circ$C) and average spring precipitation (mm) in the Northeastern United States during spring from 2020-2023.}
    \label{fig:covs}
\end{figure}

The acoustic compositional predictions in Figure \ref{fig:pred-comp} are the predicted classifications of the bioacoustic signals -- not the true signals themselves. Therefore, we make inference on bioacoustic classification, which can only serve as a proxy to the true acoustic behavior of \textit{D. pubescens}.
In general, drumming was the most commonly predicted bioacoustic behavior, whereas whinnying was the least. Whinnying was positively correlated with precipitation; see Figure \ref{fig:covs} for maps of the average spring temperature ($^\circ$C) and the average spring precipitation (mm). 
Northern New York experienced a relatively large amount of spring precipitation in 2022, resulting in higher predictive probabilities of whinnying. In 2023, northern Maine and Pennsylvania experienced less spring precipitation, resulting in lower predictive probabilities for whinnying. 
Additionally, the predictive probabilities of whinnying were affected by $\boldsymbol{\eta}$, which captured environmental effects not captured by spring temperature and precipitation.  In particular, the latent spatial effects for whinnying gave rise to slightly higher predictive probabilities in Maryland, New York, and Pennsylvania and slightly lower predictive probabilities in Massachusetts, New Hampshire, Rhode Island, and Vermont; see Appendix F for predictions of $\boldsymbol{\eta}$ across the study domain and a comparison between our proposed model and the same model without the spatial random effects $\boldsymbol{\eta}$. 
The latter model performed nearly as well as our proposed model, indicating that most of the variation in the bioacoustic compositional predictions was captured by the environmental covariates.

Drumming classification was also positively correlated with precipitation. However, the predicted spatial patterns between whinnying and drumming were not identical, suggesting that drumming was also influenced by spring temperature and had a unique latent spatial field.
Drumming was negatively correlated with spring temperature, which is most evident when comparing the northernmost and southernmost locations between whinnying and drumming classifications in Figure \ref{fig:pred-comp}.
Additionally, the latent spatial field for drumming resulted in slightly higher predictive probabilities in Maine, Maryland, Massachusetts, and Pennsylvania and slightly lower predictive probabilities in northern New York and Vermont. 
Finally, the spatial pattern of drumming classification was more consistent throughout the study period than the spatial pattern of whinnying classification, likely due to its dependence on spring temperature (which was more consistent than spring precipitation) and a smoother latent spatial field.

Piking prediction was the most variable each year, which we attribute to its negative correlation with spring precipitation. In particular, the compositional constraint required piking prediction to offset the positive correlation between whinnying/drumming and precipitation. The predictive probabilities of whinnying and drumming varied by roughly 13\% each year and had similar spatial patterns. Due to the compositional constraint, the range of piking predictive probabilities varied up to 24\% to offset the cumulative probability associated with whinnying and drumming classifications. To help satisfy the compositional constraint, the latent spatial field associated with piking classification was inversely related to the latent fields for whinnying and drumming classification.

In summary, drumming was the most commonly predicted bioacoustic signal during spring in the Northeastern United States.
The spatial heterogeneity in predictive probabilities suggests that the bioacoustic behavior of \textit{D. pubescens} was affected by the local environment. In warmer regions with high precipitation, we observed higher predictive probabilities for whinnying. In cool, wet regions, we observed an increase in predictive probabilities for drumming. In drier regions, we found higher predictive probabilities for piking.

Existing CoDA approaches are unable to obtain spatial compositional predictions, such as those depicted in Figure \ref{fig:pred-comp}, for an arbitrary compositional data set. Our model allows for positive correlations between components and zero-valued components. Directional approaches for CoDA permit these features; however, existing directional methods may not render valid compositional predictions (i.e., predicting directional points outside the non-negative orthant). Our model allows for zero-valued components, positive correlations, and is fully generative for compositional data, meaning that the predicted vector at every location in Figure \ref{fig:pred-comp} is a valid composition.

\section{Conclusion}

We developed a Bayesian hierarchical framework for spatial CoDA to analyze bioacoustic classifications. Compositional data are defined by two constraints: (i) each component must be non-negative and (ii) the sum of the components must equal a constant (often taken to be 1). However, most existing methods for CoDA assume constraints that are stronger than necessary: (i) each component must be positive and (ii) if the value of one component increases, the value of every other component must decrease (i.e., exclusively negative correlations). Our method naturally allows for zero-valued components and positive correlations between components. Therefore, our framework allows for the analysis of all compositional data without the need to manipulate or augment data (i.e., zero replacements, data imputation). Such features are appropriate when analyzing machine learning classifications, where positive correlations (between similar labels) and zero-valued components (due to confident predictions) may arise. In bioacoustic studies, passive acoustic monitoring \citep{keitt2021ecology} and deep supervised source separation \citep{sasek2024semiautomated} will increasingly render structural zeros.

Our framework features the first directional distribution truncated to the non-negative orthant, which is exclusively where transformed compositional data reside. Standard directional distributions may include support outside the non-negative orthant. In such cases, the distribution is not considered generative for compositional data because it may lead to directional observations that cannot be properly transformed back to the simplex. Existing directional approaches for compositional data either ignore this problem or circumvent it via folded distributions. Many directional distributions lack an analytically tractable normalizing constant, making folding a much more efficient approach than truncation. However, the recently proposed ESAG distribution has two key features: (i) it has an analytically tractable normalizing constant and (ii) simulation is efficient due to its relationship with the multivariate Gaussian distribution. Thus, the truncation of the ESAG distribution is computationally feasible, and we find it more intuitive than the existing folding approaches.

Compositional data are an increasingly prevalent data source in spatial statistics \citep{feng2015regression}.
However, previous directional approaches for compositional data have not accounted for spatial dependence. We proposed a spatial hyperspheric regression model that includes fixed and random spatial effects for the spatial analysis of bioacoustic signals. The latent spatial random effects were modeled as a multivariate spatial process because the relativity inherent in directional data suggests that the spatial effects for the components are interdependent. 
Although we used an LMC to model the multivariate spatial process, alternative approaches exist. Separable models are a special case of the LMC that offer computational advantages at the cost of specifying a single range parameter for all spatial fields \citep{schmidt2003bayesian}. Graphical Gaussian processes construct cross-covariance functions from graphs to ensure process-level conditional independence between variables; in highly multivariate settings, these processes offer significant computational gains over the LMC \citep{dey2022graphical}. Additionally, functional graphical Gaussian processes have been developed for multivariate functional data \citep{dey2022graphical,zapata2022partial}; \cite{chandra2024functional} developed an autoregressive matrix-Gaussian copula graphical model to infer functional connectivity in auditory systems. 
\cite{zhang2022spatial} proposed a matrix-normal approach for LMCs and spatial factor models in high-dimensional multivariate settings; \cite{alie2024computational} proposed a matrix-normal LMC that induces sparsity through the introduction of structural zeros in the coregionalization matrix.
Recently, \cite{krock2023modeling} wrote a multivariate Gaussian process as a linear combination of basis functions weighted with Gaussian graphical vectors, which scale well for highly multivariate nonstationary spatial data. Additionally, combining Hamiltonian Monte Carlo with a non-central parameterization could improve the efficiency and mixing for high-dimensional multivariate spatial processes \citep{neal2012mcmc}.

Machine learning classification provides an increasingly common source of compositional data. If the classification vectors are used for downstream analysis, then uncertainty from the classification algorithm should be propagated to the subsequent model. We allowed for uncertainty propagation in the directional analysis of compositional data in the form of $\mathbf{1}_d\otimes\boldsymbol{\alpha}'$. The uncertainty propagation term in (\ref{eq:softplus}) did not affect the relative contribution of the components nor the orientation of dispersion; rather, it solely impacted the $l_2$-norm of $\boldsymbol{\mu}$, which governed the concentration of the distribution. Thus, the auxiliary information in $\mathbf{z}_i$ was used to propagate uncertainty in our downstream model. However, many compositional data sources are the result of observation rather than prediction and, consequently, do not necessitate uncertainty propagation. For example, community composition data measure the relative proportions of species in an ecosystem and are used to infer biodiversity, monitor environmental change, and inform conservation strategies \citep{jackson1997compositional}.

While the ESAG$^+$ density is feasible to evaluate, it is more computationally expensive than its non-truncated counterpart. Future work may include multi-stage Bayesian approaches, such as recursive Bayes, to decrease the computational costs of evaluating the ESAG$^+$ density \citep[e.g.,][]{hooten2021making}. Other extensions to the present methodology might account for an autoregressive process on the regression coefficients or multivariate spatio-temporal dependencies in the data. Additionally, the concept of uncertainty propagation can be further explored in the context of compositional data, which is increasingly being generated by stochastic models.

We focused on the analysis of compositional data (i.e., bioacoustic classifications), although our methods can be adapted for unbounded directional data. For example, one only needs to set the standard ESAG distribution as the data model and use the identity link for $\boldsymbol{\mu}$. The rest of the framework remains the same to conduct a spatial directional data analysis featuring uncertainty propagation.

\section*{Acknowledgements}
We would like to thank contributors to The Macaulay Library at the Cornell Lab of Ornithology for providing the \textit{Dryobates pubescens} bioacoustic data. We also thank Nikunj Goel for helpful discussions and insights about avian acoustics. 

\section*{Funding}
This research was supported by the NSF Graduate Research Fellowship Program.

\section*{Data \& Code Availability}
\textit{We will publish our extensive code on Project Euclid upon publication. We will also provide a GitHub link containing our code.}

\bibliographystyle{agsm}
\bibliography{Compositional:Directional}

\section*{Appendix A - \textit{Dryobates pubescens} Bioacoustic Data}

We used the following recordings from the Macaulay Library at the Cornell Lab of Ornithology: 
\begin{enumerate}
    \item \textbf{Training data:} ML496665141, ML508251571, ML509560911, ML510067071, ML512308701, ML512876581,
ML513271751, ML513966371, ML514618651, ML514678191, ML516760751, ML517462401,
ML518409381, ML518754901, ML519072681, ML522725601, ML523653891, ML525553391,
ML526243831, ML528426921, ML531414841, ML532181851, ML532607181, ML533335161,
ML533682381, ML534311411, ML534396281, ML536929401, ML539778701, ML540806061,
ML540850441, ML541861321, ML542295171, ML542305721, ML543390211, ML544520671,
ML544869351, ML546095841, ML546422771, ML548045181, ML551720221, ML551726431,
ML552186051, ML552514521, ML554516391, ML554588851, ML555301551, ML555374991,
ML556412951, ML556412981, ML556611561, ML556649231, ML556704291, ML557655021,
ML557992921, ML559161651, ML560334231, ML560366301, ML560824051, ML561957241,
ML562916271, ML564833211, ML565879411, ML566486041, ML567169651, ML567768551,
ML568969091, ML569372771, ML570486411, ML572549801, ML572677821, ML575938121,
ML576349671, ML576688941, ML578716631, ML583758791, ML584530161, ML584571581,
ML585057381, ML585159931, ML585242401, ML589039401, ML591756801, ML591761031,
ML592232171, ML592245551, ML592479211, ML593473381, ML593767481, ML595784751,
ML596075211, ML597060591, ML597654171, ML597881921, ML600098931, ML600536621,
ML601691351, ML601859871, ML603921931, ML604782111, ML606584611, ML608270299,
ML608437193, ML608516473, ML608613895, ML608616919, ML608764481, ML608797524,
ML608808592, ML608858398, ML608910343, ML608912988, ML608930739, ML608951708,
ML608991727, ML609017003, ML609029053, ML609047407, ML609059779, ML609117911,
ML609133155, ML609135320, ML609135510, ML609136573, ML609137302, ML609214305,
ML609236746, ML609429636, ML609431998, ML609436518, ML609463488, ML609638039,
ML609686603, ML609834310, ML609908151, ML609918219, ML610028509, ML610144458,
ML610208639, ML610287425, ML610360923, ML610410276, ML610671802, ML610757506,
ML610811152, ML610819771, ML611100176, ML611315764, ML611315769, ML611437768,
ML611505149, ML611595837, ML611598666, ML611835566, ML611874434, ML611943821,
ML611943831, ML612041865, ML612041866, ML612168160, ML612274592, ML612276363,
ML612276516, ML612611097, ML612659335, ML612767531, ML612832449, ML612836674,
ML612880680, ML612885757, ML612959583, ML613194125, ML613384114, ML613667238,
ML613669818, ML614818215, ML615069353, ML615741436, ML615968801, ML615978417,
ML616006632, ML619667224, ML619726115
    \item \textbf{Test set:} ML212793151, ML214170691, ML215347101, ML221194451, ML227357411, ML230019701,
ML231483371, ML231585141, ML236420441, ML237150821, ML317794561, ML319441551,
ML320440621, ML320481251, ML323641781, ML323990901, ML323990911, ML323990921,
ML323990931, ML324002421, ML324002431, ML324018301, ML324022371, ML324959061,
ML324959071, ML326828571, ML327685491, ML327763201, ML327824461, ML327930381,
ML328370861, ML329660641, ML329677381, ML329679281, ML330296781, ML330542641,
ML335407641, ML336279331, ML336772261, ML343305781, ML395139451, ML423879481,
ML423987011, ML425406741, ML425739781, ML427003731, ML427006051, ML428067651,
ML428625231, ML428625881, ML428769961, ML428984271, ML428984361, ML430469361,
ML430839701, ML431921611, ML432241261, ML434825491, ML434825531, ML435034511,
ML435199601, ML435201011, ML440172101, ML442638441, ML442938041, ML443295451,
ML443295461, ML443648851, ML444425501, ML444425511, ML444543251, ML444544611,
ML446651141, ML447848371, ML447948391, ML454113781, ML459889471, ML501780791,
ML544039281, ML545150601, ML549119321, ML555466951, ML559234511, ML560380691,
ML563468531, ML565263601, ML568489991, ML568714221, ML570283871, ML570364891,
ML578818071
\end{enumerate}

\section*{Appendix B - Machine Learning Classification}

To classify the bioacoustic data, we constructed a two-dimensional convolutional neural network (2D CNN) in the TensorFlow Keras API with an input shape of (224, 224, 3) and four convolutional layers, each with a (3, 3) kernel and a rectified linear unit (ReLU) activation function 
\citep{abadi2016tensorflow,chollet2015keras}. The first convolutional layer had 32 filters, and the subsequent three layers had 128 filters. Each convolutional layer was followed by a max pooling layer with a pool size of (2, 2). Then, the two-dimensional outputs (i.e., feature maps) were flattened to a one-dimensional vector and propagated through two dense layers. The first dense layer had 1024 units, and the final dense layer had 4 units that corresponded to four categories: whinnying, drumming, piking, and noise. The noise category was included to train the architecture to distinguish between \textit{D. pubescens} signals and background noise.

To train the classification algorithm, we gathered 96 samples identified as whinnying, 28 samples identified as drumming, 36 samples identified as piking, and 23 samples identified as noise; this training data set is listed in Appendix A. The training samples were cropped to be 1-2 seconds long to isolate the appropriate signal. Then, we generated LSMSs from the cropped training data using the librosa Python package with a sample rate of 44.1 kHz \citep{mcfee2015librosa}. 
We used the LSMSs as image representations of the bioacoustic signals and as the input to the machine learning classification algorithm. The preparation and processing of the training data took approximately 20 hours.

For classification prediction, we used a sliding window approach, where a LSMS of a new recording was read in one-second intervals with half-second overlaps. For each window, we obtained a predicted probability vector. Then, we normalized the average probabilities of the three bioacoustic classes to obtain the final prediction probability vector associated with the bioacoustic signal. Finally, we squared the elements of the prediction vector to obtain the transformed directional data for downstream analysis.

\section*{Appendix C - An Alternative Parameterization}

Let $(\boldsymbol{\mu}',\boldsymbol{\gamma}')'$ be a $\mathcal{P}$-dimensional parameter vector that fully specifies $\text{ESAG}_{d-1}(\boldsymbol{\mu},\mathbf{V})$, where $\boldsymbol{\mu}=(\mu_1,\dots,\mu_d)'\in\mathbb{R}^d$ is the mean direction and $\boldsymbol{\gamma}\in\mathbb{R}^{(d-2)(d+1)/2}$ are the remaining parameters needed to specify $\mathbf{V}$. \cite{yu2024new} specified $\boldsymbol{\gamma}$ by first using a spectral decomposition of $\mathbf{V}$:
\begin{equation}
    \mathbf{V}=\sum^d_{j=1}\lambda_j\boldsymbol{\xi}_j\boldsymbol{\xi}_j',
\end{equation}
where $\lambda_j\in\mathbb{R}^+$ and $\boldsymbol{\xi}_j$ are the $j$th eigenvalue and corresponding orthonormal eigenvector of $\mathbf{V}$, respectively. According to the constraint $\mathbf{V}\boldsymbol{\mu}=\boldsymbol{\mu}$, one eigenvalue of $\mathbf{V}$ is 1 with the corresponding eigenvector $\boldsymbol{\mu}$. Thus, without loss of generality, we set $\lambda_d=1$ and $\boldsymbol{\xi}_d=\boldsymbol{\mu}/||\boldsymbol{\mu}||_2$. The derivations leading to the specification of $\boldsymbol{\gamma}$ occur in three steps: (i) parameterizing the eigenvalues $\{\lambda_j\}_{j=1}^{d-1}$, (ii) parameterizing the orthonormal eigenvectors $\{\boldsymbol{\xi}_j\}_{j=1}^{d-1}$, and (iii) grouping the eigenvalues and eigenvectors such that each group relates to elements in $\boldsymbol{\gamma}$. 

The constraint $\text{det}(\mathbf{V})=1$ implies that $\prod^{d}_{j=1}\lambda_j=1$. Given that $\lambda_d=1$, only $d-2$ parameters are needed to specify the remaining $d-1$ eigenvalues. \cite{yu2024new} introduced the radial parameters $\{r_k\}_{k=1}^{d-2}$ and specified
\begin{align}
    \lambda_1 &= \left(\prod^{d-2}_{j=1}(r_j+1)^{d-j-1}\right)^{-1/(d-1)},\\
    \lambda_j &= \lambda_1\prod^{j-1}_{k=1}(r_k+1),
\end{align}
for $j=2,\dots,d-1$. To specify the orthonormal eigenvectors of $\mathbf{V}$, we first define an orthonormal basis $(\widetilde{\boldsymbol{\xi}}_1,\dots,\widetilde{\boldsymbol{\xi}}_d)$ with $\widetilde{\boldsymbol{\xi}}_j=\mathbf{u}_j/||\mathbf{u}_j||_2$ for $j=1,\dots,d$, where 
\begin{equation}
    \mathbf{u}_j=\begin{cases}
        (-\mu_2,\mu_1,0,\dots,0)', & j=1,\\
        (\mu_1\mu_{j+1},\dots,\mu_j\mu_{j+1},-\sum_{k=1}^j\mu_k^2,0,\dots,0), & j=2,\dots,d-1,\\
        \boldsymbol{\mu}, & j=d.
    \end{cases}
\end{equation}
If $\mathbf{u}_j=\boldsymbol{0}_d$, then we set $\mathbf{u}_j=\mathbf{e}_j$, which is the unit vector with 1 at the $j$th entry.
Thus, the orthonormal basis $(\widetilde{\boldsymbol{\xi}}_1,\dots,\widetilde{\boldsymbol{\xi}}_d)$ is uniquely determined by $\boldsymbol{\mu}$. To obtain the orthonormal eigenvectors, we let $(\boldsymbol{\xi}_1,\dots,\boldsymbol{\xi}_{d-1})=(\widetilde{\boldsymbol{\xi}}_1,\dots,\widetilde{\boldsymbol{\xi}}_{d-1})\mathcal{R}_{d-1}$, where 
\begin{equation}
    \mathcal{R}_{d-1}=\left(\prod^{d-3}_{m=1}\left(R^*_{12}(\theta_{d-m-1})\prod^{d-m-2}_{j=1}R^*_{(j+1)(j+2)}(\phi_{1-j+(d-m-1)(d-m-2)/2})\right)\right)R^*_{12}(\theta_1)
\end{equation}
is a $(d-1)$-dimensional rotation matrix expressed as a product of $(d-2)(d-1)/2$ plane rotation matrices \citep{murnaghan1952element}. Each plane rotation matrix depends on a longitude angle in $[-\pi,\pi)$ or a latitude angle in $[0,\pi]$. Thus, we introduce $(d-2)(d-1)/2$ orientation parameters in the form of longitude angles $\{\theta_j\}_{j=1}^{d-2}$ and  latitude angles $\{\phi_j\}_{j=1}^{(d-2)(d-3)/2}$. The plane rotation matrix $R^*_{jk}(\cdot)$ is constructed by replacing the $(j,j)$, $(j,k)$, $(k,j)$, and $(k,k)$ elements of $\mathbf{I}_{d-1}$ with $\cos(\cdot)$, $-\sin(\cdot)$, $\sin(\cdot)$, and $\cos(\cdot)$, respectively; \cite{scealy_regression_2011} exploited this same construction to parameterize the Kent distribution. Thus, the orientation angles are introduced to parameterize the first $d-1$ eigenvectors of $\mathbf{V}$ given $\boldsymbol{\mu}$.

For $d\ge 3$, we define the set of parameters 
\begin{equation}
    \boldsymbol{\omega}=(r_1,\dots,r_{d-2},\theta_1,\dots,\theta_{d-2},\phi_1,\dots,\phi_{(d-2)(d-3)/2})',
\end{equation}
which we divide into $d-2$ groups: $(r_1,\theta_1)$ for the first group and $(r_j,\theta_j,\widetilde{\boldsymbol{\phi}}_j)$ for the $j$th group, where $\widetilde{\boldsymbol{\phi}}_j$ is the group of latitude angles of size $j-1$ for $j=2,\dots,d-2$. 
For example, if $d=5$, then $\widetilde{\boldsymbol{\phi}}_2=\{\phi_1\}$ and $\widetilde{\boldsymbol{\phi}}_3=\{\phi_2,\phi_3\}$.
We also define $\boldsymbol{\gamma}=(\widetilde{\boldsymbol{\gamma}}_1',\dots,\widetilde{\boldsymbol{\gamma}}_{d-2}')'$ as $d-2$ groups of parameters, where $\widetilde{\boldsymbol{\gamma}}_j=(\gamma_{j,1},\dots,\gamma_{j,j+1})'\in\mathbb{R}^{j+1}$. The first group is defined as $\widetilde{\boldsymbol{\gamma}}_1=(r_1\cos(\theta_1),r_1\sin(\theta_1))$, and the $j$th group is defined as 
\begin{align*}
    \gamma_{j,1} &= r_j\cos(\widetilde{\phi}_{j,1}),\\
    \gamma_{j,2} &= r_j\sin(\widetilde{\phi}_{j,1})\cos(\widetilde{\phi}_{j,2}),\\
    &\vdots\\
    \gamma_{j,j} &= r_j\sin(\widetilde{\phi}_{j,1})\sin(\widetilde{\phi}_{j,2})\dotsm\sin(\widetilde{\phi}_{j,j-1})\cos(\theta_j),\\
    \gamma_{j,j+1} &= r_j\sin(\widetilde{\phi}_{j,1})\sin(\widetilde{\phi}_{j,2})\dotsm\sin(\widetilde{\phi}_{j,j-1})\sin(\theta_j),
\end{align*}
for $j=2,\dots,d-2$. Thus, $\boldsymbol{\omega}$ fully specifies $\boldsymbol{\gamma}$. The $j$th group $\widetilde{\boldsymbol{\gamma}}_j$ can be viewed as Cartesian coordinates in the $(j+1)$-dimensional Euclidean space, which can be transformed to spherical coordinates in the $j$-dimensional spherical space occupied by the $j$th group of $\widetilde{\boldsymbol{\omega}}$. The first group of $\widetilde{\boldsymbol{\omega}}$ is $(r_1=||\widetilde{\boldsymbol{\gamma}}_1||_2,\theta_1=\text{atan2}(\gamma_{1,2},\gamma_{1,1}))$. For $j=2,\dots,d-2$, 
\begin{align*}
    r_j &= ||\widetilde{\boldsymbol{\gamma}}_j||_2,\\
    \theta_j &= \begin{cases}
        0, & \gamma^2_{j,j}+\gamma^2_{j,j+1}=0,\\
        \arccos\left(\frac{\gamma_{j,j}}{\sqrt{\gamma^2_{j,j}+\gamma^2_{j,j+1}}}\right), & \gamma_{j,j+1}\ge 0\text{ and } \gamma^2_{j,j}+\gamma^2_{j,j+1}\ne 0,\\
        -\arccos\left(\frac{\gamma_{j,j}}{\sqrt{\gamma^2_{j,j}+\gamma^2_{j,j+1}}}\right), & \gamma_{j,j+1}< 0,
    \end{cases}\\
    \widetilde{\phi}_{j,k} &= \begin{cases}
        0, & \sum^{j+1}_{l=k}\gamma^2_{j,l}=0,\\
        \arccos\left(\frac{\gamma_{j,k}}{\sum^{j+1}_{l=k}\gamma^2_{j,l}}\right), & \text{otherwise},
    \end{cases}
\end{align*}
for $k=1,\dots,j-1$. Therefore, the ESAG distribution can be parameterized by $(\boldsymbol{\mu}',\boldsymbol{\gamma}')'\in\mathbb{R}^\mathcal{P}$, which has no constraints. 

\section*{Appendix D - Simulation Study for the Normalizing Constant Approximation}
Throughout our manuscript, we analyzed 3-dimensional compositions.
In this appendix, we conduct a simulation study to select the appropriate number of Monte Carlo draws $M$ for various $d$.

In our first scenario, we specified the mean direction as 
$\boldsymbol{\mu}_d=(0.1,d^{5/2},\dots,d^{5/2})'$
for $d=\{3,5,10,15,25\}$. Such a specification for $\boldsymbol{\mu}_d$ resulted in the ESAG support overlapping a boundary of the non-negative orthant; in particular, this specification resulted in roughly the same amount of truncation for each of the specified $d$ ($\approx$50\% of the ESAG support). We simulated the elements of $\boldsymbol{\gamma}_d$ from $\text{N}(0,0.5)_0^\infty$ to ensure that the elements were non-negative. Then, we computed $\mathbf{V}_d=f(\boldsymbol{\mu}_d,\boldsymbol{\gamma}_d)$, where $f$ is defined in Appendix C. We approximated the normalizing constant of the ESAG$^+$ distribution using eq. (3) with 100 repetitions, where $M=\{100,500,1000,5000,10000,100000\}$. In Table \ref{dM}, we compared the resulting approximations with the approximation when $M=1000000$, which is taken to be the truth. 
In particular, Table \ref{dM} presents the average absolute difference of the resulting approximations with the truth. The computational costs (in seconds) of the approximations on a 3.2 GHz processor with 64 GB of RAM are presented in brackets.

\begin{table}[h]
    \centering
    \tiny
    \caption{Scenario 1: Average absolute difference between approximations and the \textit{truth}. Computational costs (in seconds) are in brackets.}
    \begin{tabular}{c|cccccc}
        \toprule
        $d$ & $100$ & $500$ & $1000$ & $5000$ & $10000$ & $100000$ \\ 
        \midrule
        $3$ & $0.0426$ [$3.89$e-5] & $0.0200$ [$1.33$e-4] & $0.0122$ [$2.67$e-4] & $0.0054$ [$0.0014$] & $0.0036$ [$0.0027$] & $0.0014$ [$0.0296$] \\
        $5$ & $0.0439$ [$5.67$e-5] & $0.0179$ [$2.00$e-4] & $0.0149$ [$4.00$e-4] & $0.0056$ [$0.0020$] & $0.0042$ [$0.0040$] & $0.0016$ [$0.0402$] \\
        $10$ & $0.0400$ [$8.28$e-5] & $0.0176$ [$2.94$e-4] & $0.0136$ [$5.79$e-4] & $0.0069$ [$0.0029$] & $0.0042$ [$0.0058$] & $0.0014$ [$0.0677$] \\
        $15$ & $0.0404$ [$9.40$e-5] & $0.0180$ [$3.61$e-4] & $0.0123$ [$7.06$e-4] & $0.0052$ [$0.0036$] & $0.0044$ [$0.0072$] & $0.0014$ [$0.0807$] \\
        $25$ & $0.0401$ [$1.64$e-4] & $0.0183$ [$6.66$e-4] & $0.0125$ [$1.32$e-3] & $0.0055$ [$0.0066$] & $0.0041$ [$0.0134$] & $0.0014$ [$0.1515$] \\
        \bottomrule
    \end{tabular}
    \label{dM}
\end{table}

Our second scenario only differed from the first scenario in the specification of $\boldsymbol{\mu}$: $\boldsymbol{\mu}_d=(3/2,d^5,\dots,d^5)'$, which resulted in more truncation for larger $d$. In particular, 95\% of the ESAG$_3$ support was in the non-negative orthant, while only 51\% of the ESAG$_{25}$ support was in the non-negative orthant. Table \ref{dM2} reveals that the approximation is less accurate for larger $d$ (i.e., more truncation). Computational costs are excluded from Table \ref{dM2} because they are roughly equivalent to the costs in Table \ref{dM}.

\begin{table}[h]
    \centering
    \small
    \caption{Scenario 2: Average absolute difference between approximations and the \textit{truth}.}
    \begin{tabular}{c|cccccc}
        \toprule
        $d$ & $100$ & $500$ & $1000$ & $5000$ & $10000$ & $100000$ \\ 
        \midrule
        $3$ & $0.0193$ & $0.0074$ & $0.0057$ & $0.0022$ & $0.0014$ & $0.0006$ \\
        $5$ & $0.0210$ & $0.0102$ & $0.0062$ & $0.0036$ & $0.0026$ & $0.0008$ \\
        $10$ & $0.0301$ & $0.0139$ & $0.0108$  & $0.0044$ & $0.0029$ & $0.0011$ \\
        $15$ & $0.0371$ & $0.0185$ & $0.0115$  & $0.0059$ & $0.0036$ & $0.0013$ \\
        $25$ & $0.0400$ & $0.0190$ & $0.0120$ & $0.0067$ & $0.0039$ & $0.0014$ \\
        \bottomrule
    \end{tabular}
    \label{dM2}
\end{table}

In our applications, we set $M=5000$ because it renders the quickest approximation with an average discrepancy that falls within an acceptable tolerance of $10^{-2}$ for both scenarios.
Additionally for each MCMC iteration, we approximate the integral $n(d+1)^2/2 + nE_k$ times, where $E_k$ is the number of iterations in the elliptical slice sampler (ESS) for $\boldsymbol{\eta}$ on the $k$th MCMC iteration. The average number of ESS iterations per MCMC iteration is 12.1 and 9.4 for the simulation study and bioacoustic study, respectively. Therefore, the computational savings of smaller $M$ scale quadratically with $d$.

\section*{Appendix E - The Complete Bayesian Hierarchical Model}

We fit the following Bayesian hierarchical model in our simulation and bioacoustic case studies:
\begin{align*}
    [\mathbf{y}_i\mid\boldsymbol{\mu}_i,\mathbf{V}_i]^+ &= \frac{[\mathbf{y}_i\mid\boldsymbol{\mu}_i,\mathbf{V}_i]\mathbbm{1}(\mathbf{y}_i\in\mathcal{S}^{d-1}_+)}{\int_{\mathcal{S}^{d-1}_+} [\mathbf{y}_i\mid\boldsymbol{\mu}_i,\mathbf{V}_i]d\mathbf{y}_i},\\
    [\mathbf{y}_i\mid\boldsymbol{\mu}_i,\mathbf{V}_i] &= \text{ESAG}_{d-1}(\boldsymbol{\mu}_i,\mathbf{V}_i),\\
    \boldsymbol{\mu}_i &= \text{softplus}(\mathbf{Bx}_i+\boldsymbol{\eta}(\mathbf{s}_i) + \mathbf{A}\mathbf{z}_i),\\
    \mathbf{V}_i &= f(\boldsymbol{\mu}_i,\boldsymbol{\gamma}), \\
    \boldsymbol{\beta}_j &\sim \text{N}_p(\boldsymbol{0},\sigma^2_\beta\mathbf{I}),\\
    \mathbf{H} &\sim \text{LMC}(\mathbf{C},\mathbf{R}^d_{j=1}),\\
    \boldsymbol{\alpha} &\sim \text{N}_q(\boldsymbol{0},\sigma^2_\alpha\mathbf{I}),\\
    \gamma_{l,k} &\sim \text{N}(0,\sigma^2_\gamma),\\
    C_{ij} &\sim \text{N}(0,1),\\
    \phi_j &\sim \text{Gamma}(\alpha_\phi,\theta_\phi),
\end{align*}
where $i=1,\dots,n$, $j=1,\dots,d$, $l=1,\dots,d-2$, and $k=1,\dots,l+1$. The hyperparameters are specified for the simulation and bioacoustic case studies in their corresponding sections. The function $f$ to compute $\mathbf{V}$ deterministically is described in Appendix C.

\section*{Appendix F - Latent Spatial Random Effects}

\begin{figure}
    \centering
    \includegraphics[width=\textwidth]{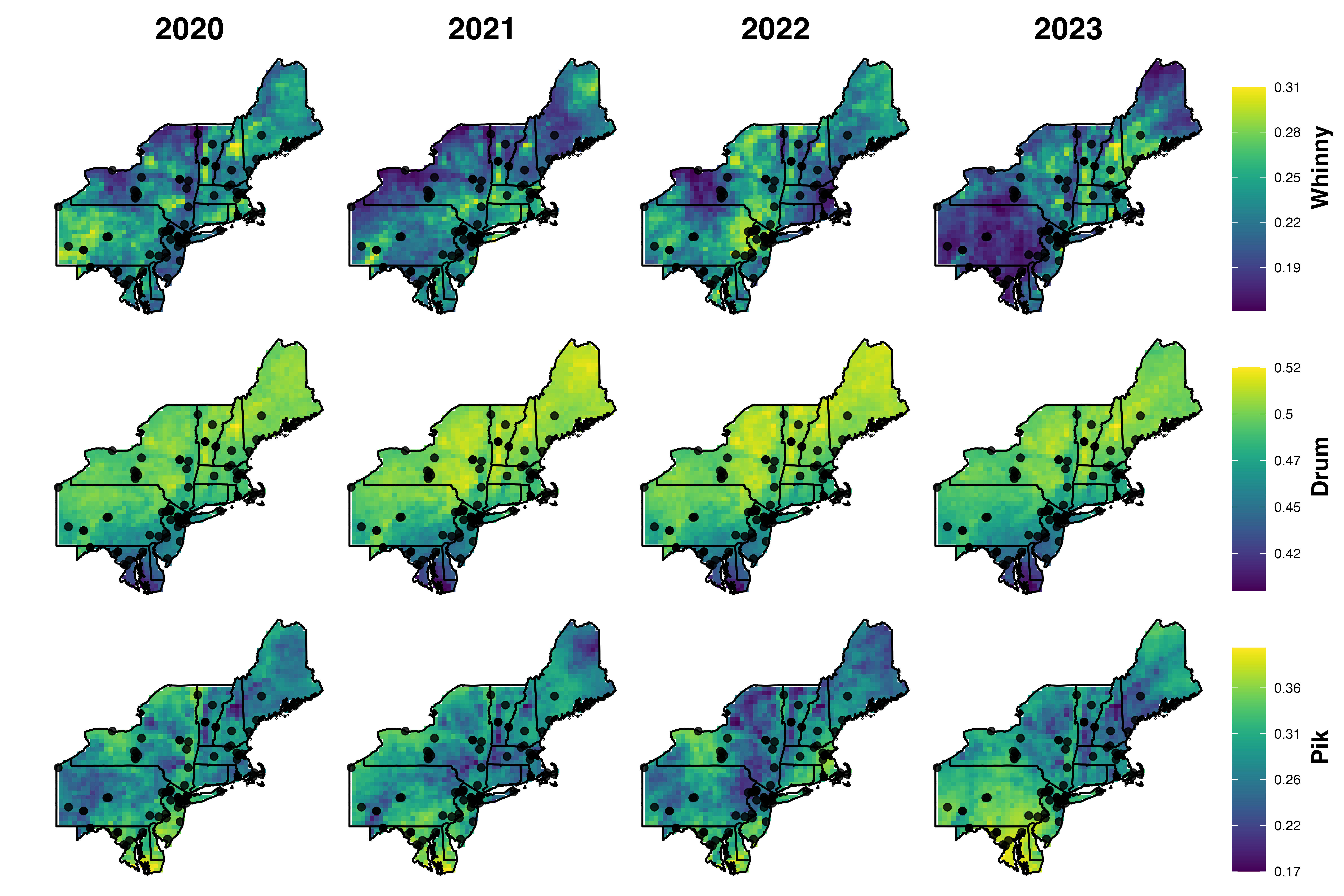}
    \caption{Predicted \textit{Dryobates pubescens} bioacoustic classifications for the Northeastern United States using the non-spatial model. Black dots denote observed locations.}
    \label{fig:nonspatial}
\end{figure}

We compared the model in Appendix E with the same model excluding the latent spatial random fields $\boldsymbol{\eta}$; we refer to the former as the spatial model and the latter as the non-spatial model, although both contain environmental covariates that are spatially defined. We fit the non-spatial model to the square-root transformed \textit{D. pubescens} compositional predictions detailed in Section 2. In the non-spatial model, we used the same priors and hyperparameters as specified in Section 7 and ran the MCMC algorithm for 400000 iterations (as we did with the spatial model). Then, we obtained spatial compositional predictions using the method outlined in Section 7.1, where $\boldsymbol{\eta}^{(k)}(\mathbf{s}^*)= 0$ for every location $\mathbf{s}^*\in\mathcal{S}$ and MCMC iteration. The resulting spatial compositional predictions are depicted in Figure \ref{fig:nonspatial}.

\begin{figure}
    \centering
    \includegraphics[width=\textwidth]{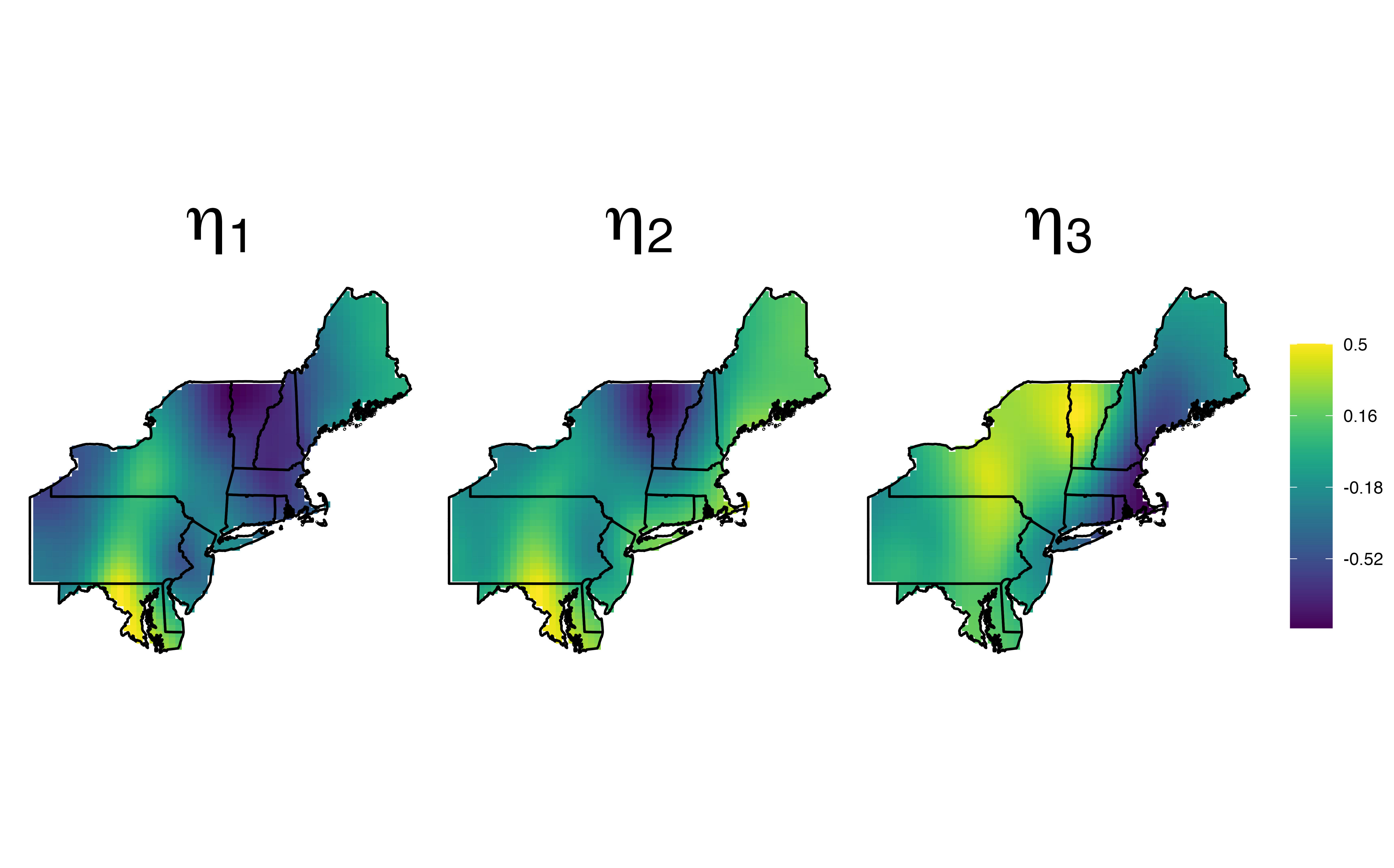}
    \caption{Predicted latent spatial random fields $\boldsymbol{\eta}$ for bioacoustic signal classifications of \textit{D. pubescens}.}
    \label{fig:eta}
\end{figure}

The predictive probabilities for whinnying and piking classification were similar between the spatial and non-spatial models. Although there were fine-scale differences, both models captured the same positive (negative) relationship between precipitation and whinnying (piking). However, the drumming predictive probabilities were significantly different. 
In the spatial model, drumming classification was positively correlated with precipitation and negatively correlated with temperature; the latent spatial field for drumming (depicted in Figure \ref{fig:eta}) shared several features with average spring temperature. 
In the non-spatial model, predicted drumming classifications were dominated by the negative correlation with temperature.

Drumming (the baseline category) was the only component that significantly differed.
Due to the truncation of the ESAG to the non-negative orthant, we defined a baseline category (drumming) and set its regression coefficients to zero to induce identifiability. No such constraint is needed for the multivariate spatial process $\boldsymbol{\eta}$. The latent spatial effects impact both the location parameter $\boldsymbol{\mu}$ and the concentration of the data model (via $||\boldsymbol{\mu}||_2$). Therefore, when excluding $\boldsymbol{\eta}$ from the model, we lost power in predicting the baseline component. As a result, the main discrepancy between the spatial and non-spatial model output was the predictive probabilities for drumming.


Finally, we used the logarithmic score function (logS) for quantitative model comparison \citep{bjerregaard2021introduction}. The mean logS were 2.0452 and 2.1631 for the spatial and non-spatial model, respectively, indicating that the spatial model provided a slightly better fit to the data.

We also fit the spatial and non-spatial model to two simulated data sets. Both data sets were simulated using the same approach as in Section 6 with $\boldsymbol{\phi}=(0.12, 0.15, 0.19)'$, $\alpha=1.0494$, $\boldsymbol{\gamma}=(0.9554,1.3055)'$, 
$$\mathbf{B}=\begin{bmatrix}
1.0 & 0.8 & 0.5 \\
1.25 & 1.15 & 0.25 \\
1.1 & 0.55 & 1.65 
\end{bmatrix} \; , \; \text{  and } \; \mathbf{C}=\begin{bmatrix}
-1.71 & -2.59 & 1.21 \\
0.63 & 0.07 & 0.94 \\
-0.33 & -0.24 & -1.55 
\end{bmatrix}.$$ As in Section 6, the covariates for each observation were fully specified by their spatial location: $\mathbf{x}(\mathbf{s}_i)=(1,x(\mathbf{s}_i)_1,x(\mathbf{s}_i)_2)'$ for $i=1,\dots,100$, where $x(\mathbf{s}_i)_1=|s_{i,1}-0.5|^{1.2}$ and $x(\mathbf{s}_i)_2=||\mathbf{s}_i||_2$.
The first simulated data set contained spatial random effects $\boldsymbol{\eta}$, whereas we fixed $\boldsymbol{\eta}=\boldsymbol{0}$ in the second simulated data set (i.e., the compositional responses were fully specified by the environmental covariates). We compared the spatial and non-spatial models under these two simulated scenarios to showcase the flexibility of the multivariate spatial process.

\begin{figure}
    \centering
    \includegraphics[width=\textwidth]{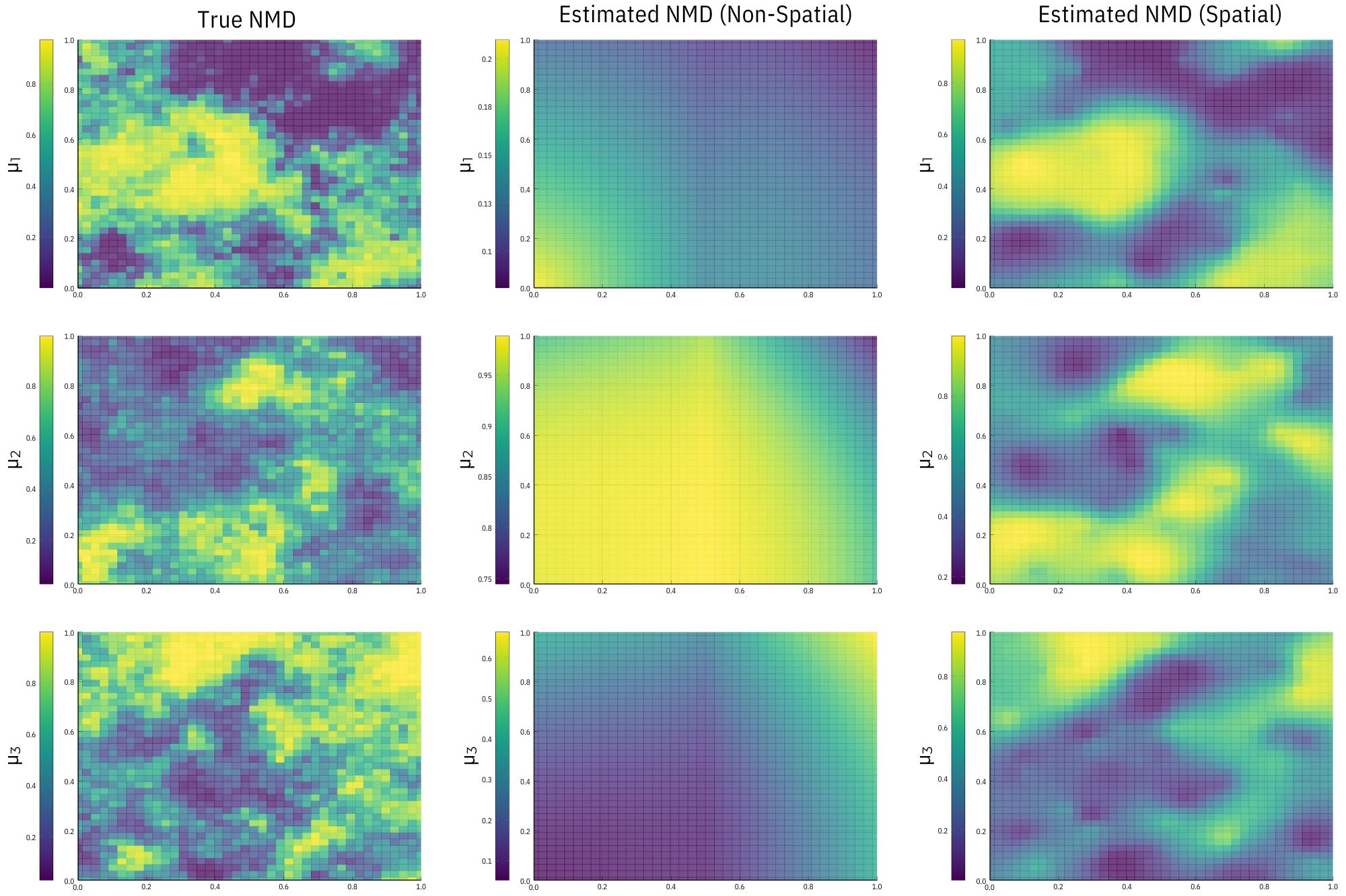}
    \caption{The true normalized mean direction (NMD) and estimated NMD for the spatial and non-spatial models under the first simulated data set.}
    \label{fig:nmd_spatial}
\end{figure}

In the first scenario, the spatial model ($\text{logS}=0.6931$) significantly outperformed the non-spatial model ($\text{logS}=1.8863$). Figure \ref{fig:nmd_spatial} displays the true normalized mean direction (NMD) and the NMD estimated by each model. The spatial model provided a much better fit to the data because the compositions were simulated with latent spatial fields. In our bioacoustic case study, most of the variation in the compositional predictions was explained by the environmental covariates. However, in other applications, the available environmental covariates may not fully explain the spatial structure of the compositional response. In such cases, the latent spatial fields can significantly improve the model fit, as exemplified in this model comparison.

\begin{figure}
    \centering
    \includegraphics[width=\textwidth]{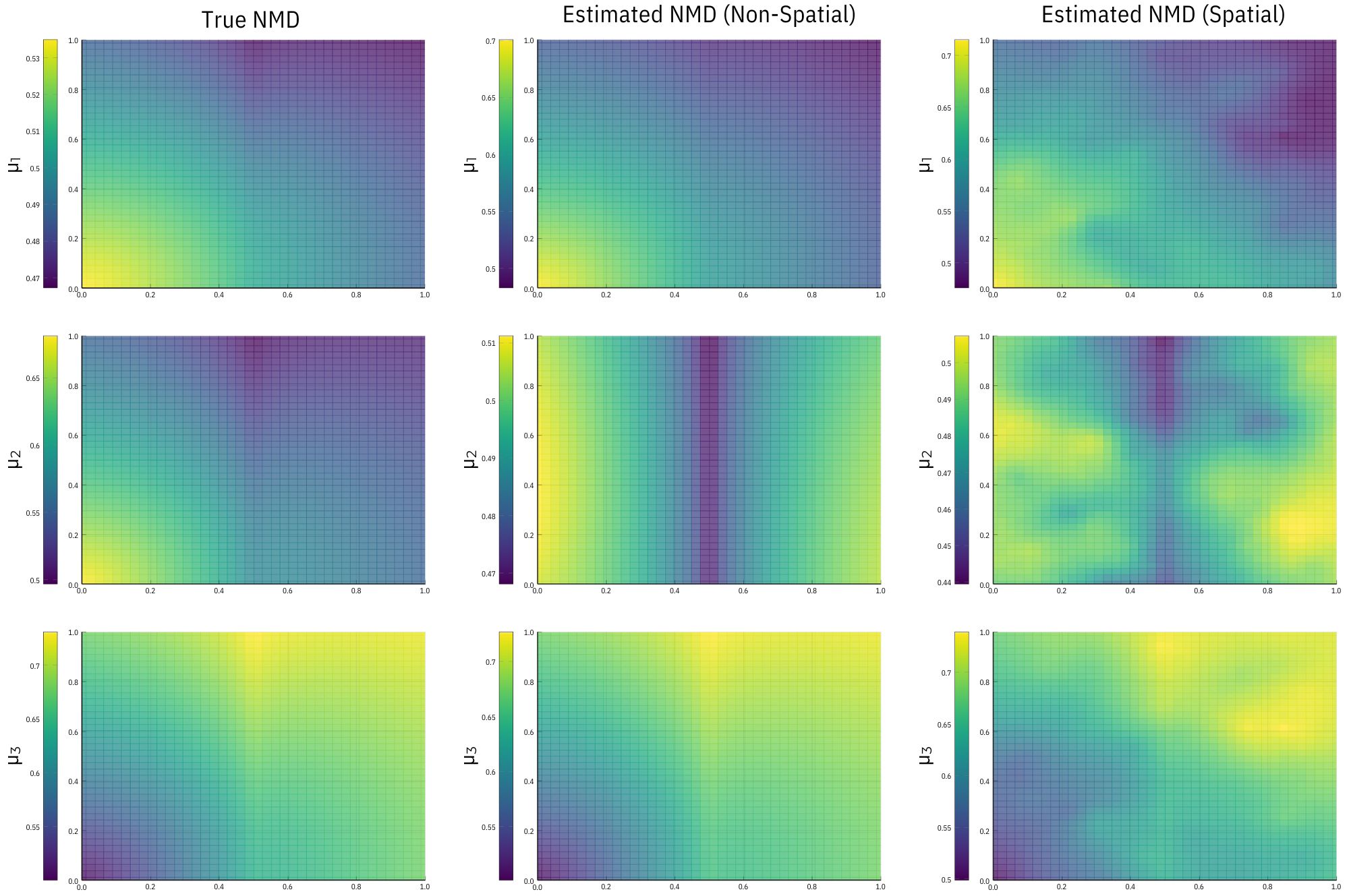}
    \caption{The true normalized mean direction (NMD) and estimated NMD for the spatial and non-spatial models under the second simulated data set.}
    \label{fig:nmd_nonspatial}
\end{figure}

In the second scenario, the non-spatial model ($\text{logS}=-1.3627$) only slightly outperformed the spatial model ($\text{logS}=-1.1264$). As shown in Figure \ref{fig:nmd_nonspatial}, both models adequately captured the first and third components yet overestimated the magnitude of $B_{2,3}$. 
Although the spatial model predicted a noisier NMD surface, the overall pattern was similar to the surface predicted by the non-spatial model. Thus, the latent spatial fields $\boldsymbol{\eta}$ can account for spatial heterogeneity not captured by environmental covariates (as seen in the first simulated model comparison) or they can have a minimal impact on inference (as seen in the second simulated model comparison).

\end{document}